\documentclass[english,fleqn]{article}
\usepackage{ae,aecompl}
\usepackage[T1]{fontenc}
\usepackage[latin9]{inputenc}
\usepackage[a4paper]{geometry}
\usepackage{color}
\usepackage{amsmath}
\usepackage{amssymb}
\usepackage{setspace}
\usepackage{esint}
\usepackage{graphicx}

\makeatletter
\newcommand{\address}[1]{
\par {\raggedright #1
\vspace{1.4em}
\noindent\par}
}

\makeatother

\usepackage{babel}
\def\be{\begin{equation}}
\def\ee{\end{equation}}
\def\bea{\begin{eqnarray}}
\def\eea{\end{eqnarray}}

\def\nn{\nonumber}

\def\phicirc{\stackrel{\,\,\circ}{\phi}}
\def\phibullet{\stackrel{\,\,\bullet}{\phi}}
\def\psicirc{\stackrel{\,\,\circ}{\psi}}
\def\psibullet{\stackrel{\,\,\bullet}{\psi}}

\def\phibarbullet{\stackrel{\,\,\bullet}{\bar\phi}}

\def\Hcirc{ {\cal H}_{\rm m}^{(\circ)} }
\def\Hbullet{{\cal H}_{\rm m}^{(\bullet)}}
\def\Vcirc{ {\cal V}_{\rm m}^{(\circ)} }
\def\Vbullet{{\cal V}_{\rm m}^{(\bullet)}}
\def\Pcirc{ P^{(\circ)} }
\def\Pbullet{P^{(\bullet)}}

\def\Hc{{\cal H}_{\rm c}}

\begin{document}

\title{\textbf{$SU(2)$-particle sigma model:  Momentum-space quantization of a particle on the sphere $S^3$} }

\author{J. Guerrero$^{a}$, F.
F. L\'opez-Ruiz$^b$ and V. Aldaya$^{c}$}

\maketitle
\begin{singlespace}

\address{\noindent \begin{center}
\emph{$^{a}$Departamento de Matem\'aticas, Universidad de Ja\'en, Campus las Lagunillas, 23071 Ja\'en, Spain}\\
\emph{$^{b}$Departamento de F\'\i sica Aplicada, Universidad de C\'adiz,
Campus de Puerto Real, E-11510 Puerto Real, C\'adiz, Spain}\\
\emph{$^{c}$Instituto de Astrof\'\i sica de Andaluc\'\i a (IAA-CSIC), Glorieta
de la Astronom\'\i a, E-18080 Granada, Spain}\\

\emph{}\\
\emph{julio.guerrero@ujaen.es~~ paco.lopezruiz@uca.es~~ valdaya@iaa.es}
\par\end{center}}
\end{singlespace}
\begin{abstract}

We perform the momentum-space quantization of a spin-less particle moving on the $SU(2)$ group
manifold, that is, the three-dimensional sphere $S^{3}$, by using a non-canonical method
entirely based on symmetry grounds. To achieve this task, non-standard (contact) symmetries
are required as already shown in a previous article where the configuration-space quantization was given. 
The Hilbert space in the momentum space representation turns out to be made of a subset of (oscillatory) solutions of the Helmholtz equation in four dimensions.
The most relevant result is the fact that both the scalar
product and the generalized Fourier transform between configuration and momentum spaces deviate notably from the naively expected expressions, the former exhibiting now a non-trivial
kernel, under a double integral, traced back to the non-trivial topology of the phase space, even
though the momentum space as such is flat. In addition, momentum space itself appears directly as the carrier space of an irreducible representation of the symmetry group,
and the Fourier transform as the unitary equivalence between two unitary irreducible representations.

%

\end{abstract}

\section{Introduction}

The quantization of a spin-less particle moving on the $SU(2)$ group
manifold, that is, the $S^3$ sphere, was achieved in a previous
paper \cite{SU(2)} by resorting  to a non-canonical,
group-theoretical algorithm, there referenced, fully appropriate
for non-linear systems and/or bearing a non-trivial topology, where
Canonical Quantization turns out not to be adequate. This symmetry
based quantization procedure required the use of symmetry
transformations which are not point symmetries of the classical Lagrangian,
but (contact) symmetries of the associated Poincar\'e-Cartan
form, which expand the full phase space of the system. These symmetries close the $\tilde\Sigma(SU(2))$ group (see Sec. \ref{Configuracion}), which is a proper subgroup of the Euclidean $E(4)$ group\footnote{For a spinning particle the symmetries of the Poincar\'e-Cartan form close the full $E(4)$ group. In general, for a particle moving in $S^n$ with $n\neq 3$ the symmetries close the group $E(n+1)$.}. 
See \cite{Gadella} for other algebraic approaches for the same problem (see also \cite{Santander1,Santander2} for the case of $S^2$).

We chose there the more direct ``representation'' provided by the
group approach in this example, which is the configuration-space
one, where wave functions depend on co-ordinates on the sphere. As expected, the algorithm exhibits a proper and unambiguous
realization of the basic operators and of the Hamiltonian (which turns out to be the Laplace-Beltrami operator on $S^3$) as well as a non-trivial
integration measure in the scalar product naturally attached to
the topology of the sphere. Those results fit alternative approaches that can be found in the literature for $S^2$ \cite{Liu,RMP}.

 Naturally, the question arises of whether some sort of 
``momentum'' representation exists in such system bearing a configuration space
with compact topology (without boundary), in which wave functions would depend on 
``tangent velocities'' or ``momenta'', in analogy with the free Galilean particle. Traditionally, the momentum space representation is related to
the configuration space one through the Fourier Transform, with its mathematical foundations in the Pontryagin duality theory \cite{Folland}, where the momentum space is the Pontryagin dual (i.e. the set of unitary  characters of the Abelian group constituted by the translations in configuration space). When the configuration space is a compact Abelian group, the associated
Pontryagin dual is discrete, and the Fourier transform reduces to the Fourier Series (see \cite{GeneralPhaseSpaces} for a review on the different cases for Abelian configuration spaces). When the configuration space is a non-Abelian group, Pontryagin duality theory becomes more complicated, and 
the associated Pontryagin dual is made of unitary and irreducible representations of different (even infinite) dimensions, rendering the Fourier transform cumbersome and the interpretation of momentum space
as a manifold unclear (the Pontryagin dual can even become a non-Hausdorff manifold \cite{Folland}).
An alternative and simpler description of momentum space for non-Abelian groups, which does not rely on Pontryagin duality theory and parallels the Abelian case, is given by the Sherman-Volobuyev construction \cite{Sherman,Volobuyev}. There, an overcomplete (and non-orthogonal) basis  in configuration space and its dual are given, whose labels, both discrete and continuous, play the role of a pair of dual ``momentum spaces'' \cite{WolfWignerFunction-Sphere}.

However, a different description is possible, where the momentum space appears as the continuous manifold supporting the carrier space of an irreducible and unitary representation of the minimal group of symmetries of the Poincar\'e-Cartan 1-form, $\tilde\Sigma(SU(2))$, and the Fourier
transform as the unitary equivalence between this representation and the previously given on the configuration space \cite{SU(2)}. Both representations are obtained, by using our group-theoretical algorithm,  with two different (though equivalent) \textit{polarizations} (see below).

In the present article, we face the momentum-space quantization of a particle moving on $S^3$ and we shall discover that the new
``representation'' proves to be quite non-trivial, even though the
momentum space itself is flat. In fact, given that momentum space
is the (co)-tangent space of the $S^3$ space at a point, one might
expect that a momentum-space ``representation'' would behave
similarly to the flat, free case. But, on the other hand, it is
clear that the non-trivial topology of the configuration space
should show up somehow irrespective of the quantum
``representation''. Indeed, the scalar product and
accordingly the generalized Fourier transform relating both
``representations'' deviate from the standard form. In particular,
the carrier space of the representation in momentum space is made out of the subset of \textit{oscillatory} solutions of the
Helmholtz equation 
and the scalar product naturally appears involving a double
integration with a kernel based on Bessel functions instead of
Dirac's deltas. This kind of scalar product is not new and was
introduced in particular in Helmholtz Optics \cite{BernardoOptica}, 
though interchanging the role of momenta and co-ordinates.

Our construction of the representation in momentum space is derived intrinsically from the basic symmetry group $\tilde\Sigma(SU(2))$, and Helmholtz equation appears naturally as the eigenvalue equation for the Casimir of $\tilde\Sigma(SU(2))$ in the momentum representation.
Although surprising, the fact that Helmholtz equation should appear when we try to construct a momentum representation for the particle on the sphere is rather intuitive if we use any approach to quantization using constraints, since the  constraint $x_1^2+x_2^2+x_3^2+x_4^2-R^2= 0$
at the quantum level, expressed  in the momentum representation instead of the usual configuration representation, i.e. with $\hat{x}_i=i\hbar \frac{\partial\,}{\partial p_i}$, leads to Helmholtz equation.



In this paper we very briefly present in Sec. \ref{Configuracion}, as a reminder,
the framework for the quantization of the particle moving on the
sphere $S^3$, in a way that serves at the same time as an
illustration of the group approach to quantization method and for fixing the notation. 
In Sec. \ref{Momento} we face the central subject of the quantization on
momentum space, which requires the introduction of a higher-order
polarization restriction leading to Helmholtz equation, and a proper scalar product. In Sec. \ref{Fourier} the unitary
correspondence between configuration and momentum space
``representation'' is established as a generalized Fourier
Transform, allowing for a new resolution of the identity in momentum space, which reveals a reproducing kernel Hilbert space structure. 
In Section \ref{TimeEvolutionMomentum} the Hamiltonian describing the free spin-less particle on $SU(2)$ is introduced in momentum space and time evolution is studied.
 We finally conclude with some comments on future
applications and or extensions of the present research. Several Appendices are given at the end with some technical material; we shall also
take the opportunity of completing some further details which
where absent from the basic study in the previous paper \cite{SU(2)}.

\section{Brief report on group-theoretical quantization}
\label{Configuracion}

The classical motion of a particle of mass $m$ on the $SU(2)$-group manifold corresponds to that of a Particle Non-Linear Sigma Model on the 
Riemannian manifold $S^3$, with Lagrangian obeying the standard form:
\begin{equation}
L=\frac{1}{2}m g_{ij}\dot{\epsilon}^{i}\dot{\epsilon}^{j}=\frac{1}{2} m\,k_{ab}\theta_{i}^{R\,(a)}\theta_{j}^{R\,(b)}\dot{\epsilon}^{i}\dot{\epsilon}^{j}\;
=\;\{R\rightarrow L\},\label{Lagrangian}
\end{equation}
where $\vec\epsilon=(\epsilon^1,\epsilon^{2},\epsilon^3)\in B_R\equiv B(\vec 0,R)=\{\vec x\in\mathbb{R}^3\,,\|\vec x\|<R\}$ are local co-ordinates of the 
$SU(2)$-group ($\dot{\epsilon}^{i}=\frac{d\epsilon^{i}}{dt}$),  
$k_{ab}=\delta_{ab}$ is the Cartan-Killing metric, and $\theta_{i}^{R(a)}$ the components of the right-invariant 1-forms of Cartan for the $SU(2)$ group.
Parameterizing locally the group elements 
as rotations in three dimensions  along the axis $\vec\epsilon$ through an angle $\varphi=2\hbox{sin}^{-1}\frac{\|\vec\epsilon\|}{R}$, the group manifold 
can be identified with the points of a sphere of radius $R$ embedded in four dimensions, $S^3$, where the fourth component $\epsilon_4$ verifies ${\vec\epsilon}{\,}^2+\epsilon_4^2=R^2$, and therefore 
$|\epsilon_4|\equiv R\rho(\vec\epsilon)=R \sqrt{1-\frac{\vec{\epsilon\,}^{2}}{R^2}}$. Alternatively, each point in $S^3$ can be parametrized by hyperspherical coordinates, related
 to the coordinates $({\vec\epsilon},\epsilon_4)$ through
\bea
\epsilon_1&=&R\,\sin{\chi}\,\sin{\theta}\,\cos{\phi}\nn\\
\epsilon_2&=&R\,\sin{\chi}\,\sin{\theta}\,\sin{\phi} \label{hypersphericalcoor}\\
\epsilon_3&=&R\,\sin{\chi}\,\cos{\theta}\nn\\
\epsilon_4&=&R \cos{\chi}\,.\nn
\eea

\begin{figure}[h]
\begin{center}
\includegraphics[height=4cm]{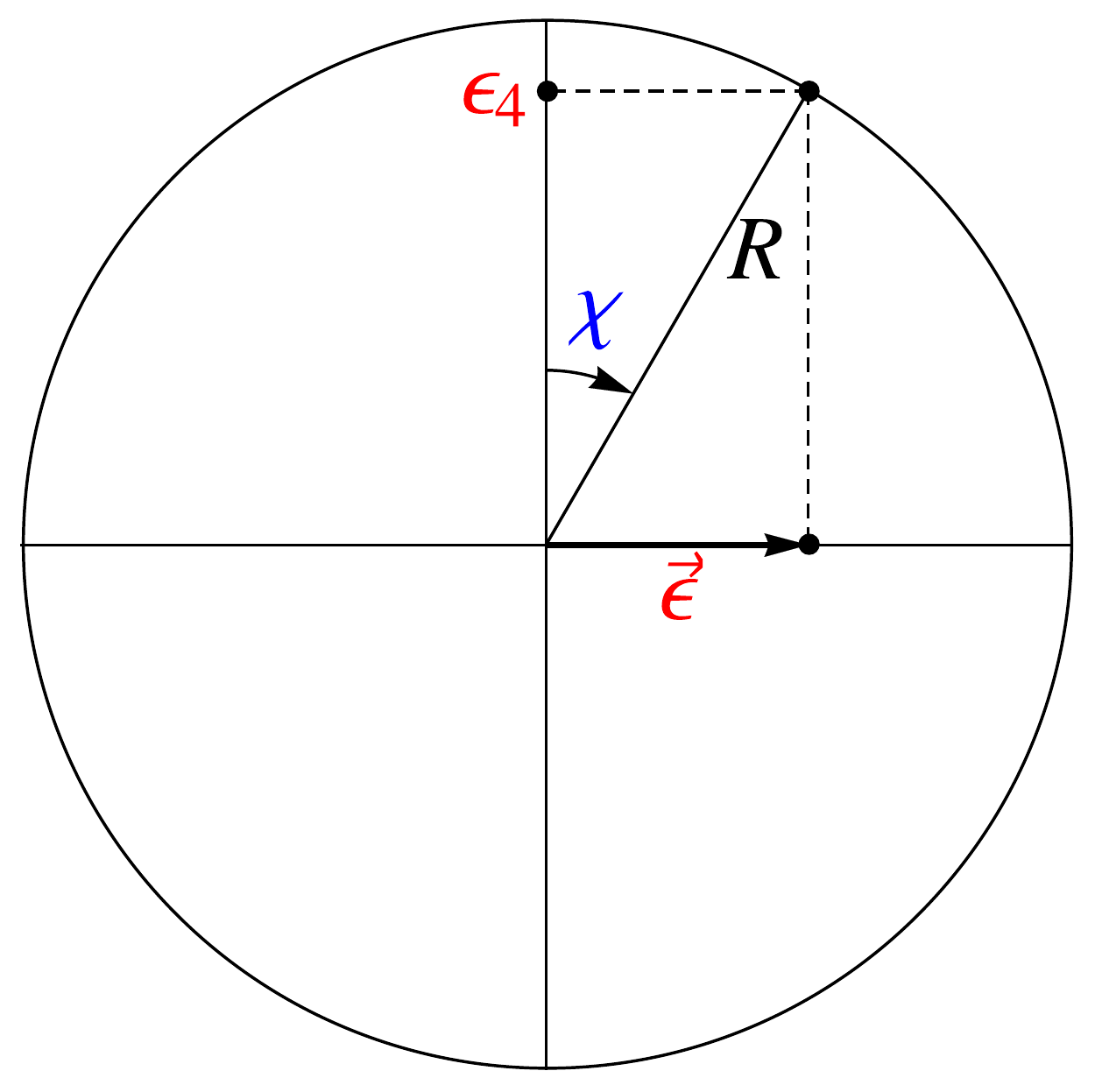}   \hskip 2cm \includegraphics[height=4cm]{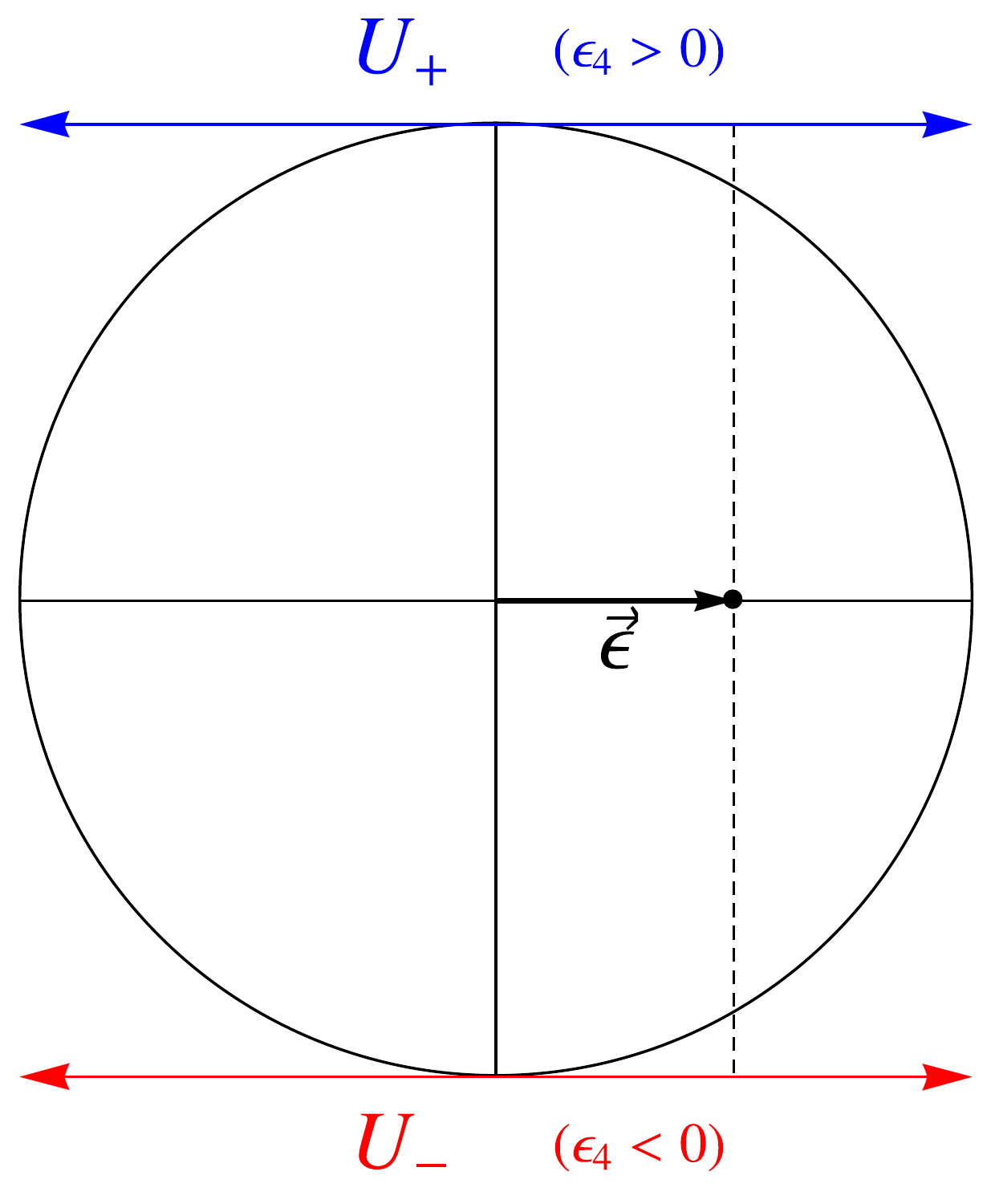} \hskip 2cm
\includegraphics[height=4cm]{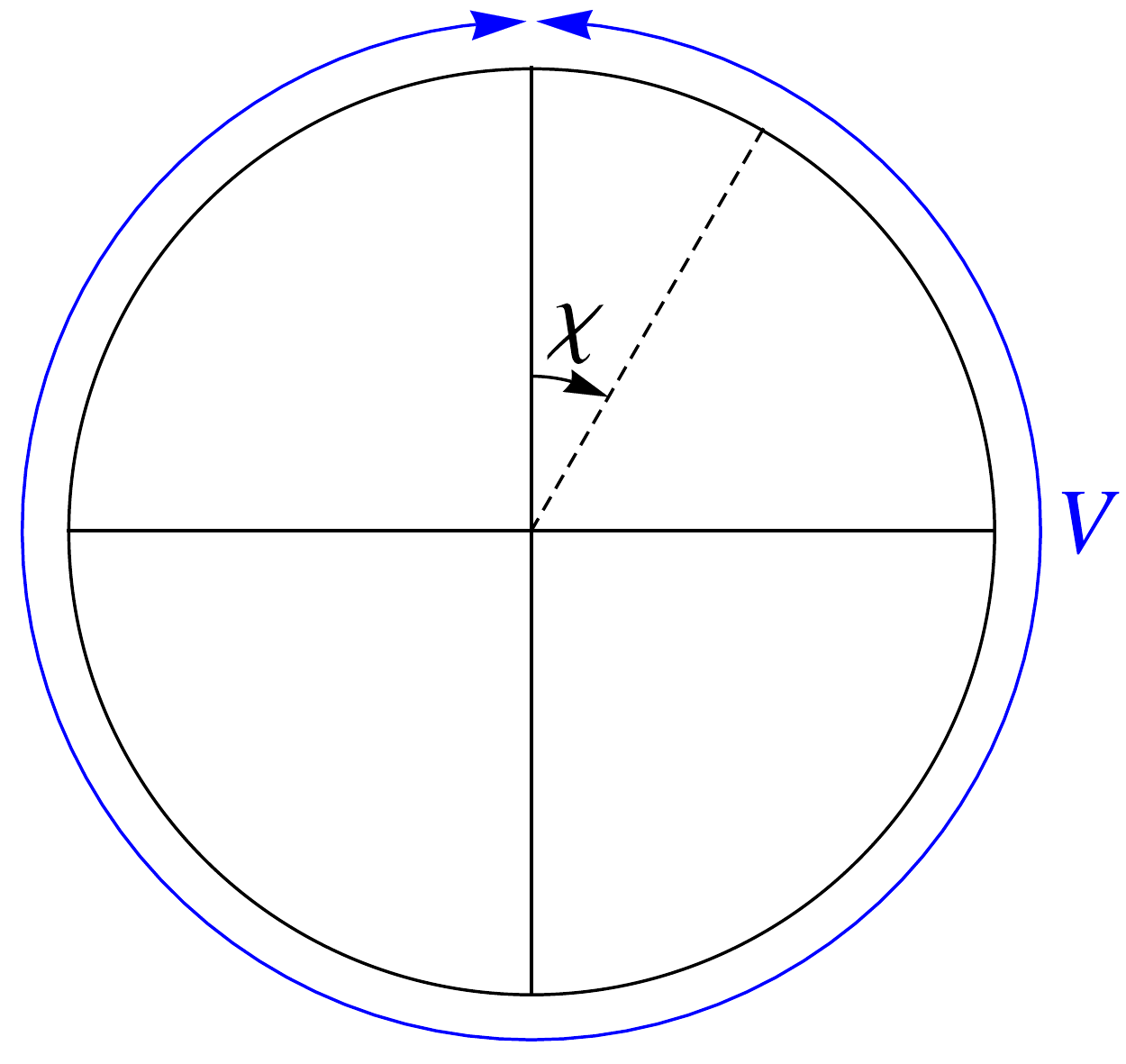}
\end{center}
\caption{(Left) Relation between cartesian coordinates $(\vec{\epsilon},\epsilon_4)$ and hyperspherical coordinates $(\chi,\theta,\phi)$. (Center) Cartesian coordinate charts. (Right) Hyperspherical coordinate chart.
For all cases, a cut through the hyperplane $\epsilon_1=0,\epsilon_2=0$ ($\theta=0,\phi=0$) is shown. \label{Fig1}}
\end{figure}

 See Fig. \ref{Fig1} for a representation of (a section of) these coordinates on the sphere $S^3$. Note that we require two charts, $U_+$ and $U_-$, to cover $S^3$ with the coordinates $\vec{\epsilon}$ (up to a set of zero measure, the equatorial sphere $S^2$, also of radius $R$). 
 In the following, we shall mainly work with the local charts $U_+$ and $U_-$, although the local expressions will be given in the chart $U_+$, the expressions on the chart $U_-$ are easily obtained by changing the sign to $\epsilon_4$. The local chart $V$ will be used when the symmetry of the involved objects simplifies the expressions in hyperspherical coordinates.

With these co-ordinates, the ``square root of the metric'' or vierbeins (in the present example the, let us say, right-invariant 
canonical $1$-forms on the group) acquire the expression:
\begin{equation}
\theta^{R\,(i)}=\theta_{j}^{R\,(i)}d\epsilon^{j}=\frac{1}{R}\left(\pm|\epsilon_4|\delta_{j}^{i}\pm
\frac{\epsilon^{i}\epsilon_{j}}{|\epsilon_4|}+\eta_{\cdot jk}^{i}\epsilon^{k}\right)\, d\epsilon^{j},
\end{equation}
so that the Lagrangian becomes:
\begin{equation}
L=\frac{1}{2}m\,\left(\delta_{ij}\pm
\frac{\epsilon_{i}\epsilon_{j}}{|\epsilon_4|^2}\right)\dot{\epsilon}^{i}\dot{\epsilon}^{j}\,.\label{Lagrangiano}
\end{equation}
The corresponding Euler-Lagrange equations of motion can be cast in the simple form  
\begin{equation}
 \frac{\!d}{dt}(\theta^i\equiv \theta^{R\,(i)}_k\dot{\epsilon^k})=0\,\,\,\longleftrightarrow\,\,\,\epsilon^{i}=\varepsilon^{i}\cos\omega t+
  \dot{\varepsilon}^{i}\frac{\sin\omega t}{\omega}\,,
\end{equation}
where the constants $\varepsilon^i$ and $\dot\varepsilon^i$ are the initial values of co-ordinates and velocities, 
$\omega\equiv\sqrt{\frac{8}{mR^2}H}=
\frac{1}{R}\sqrt{g_{ij}(\vec{\varepsilon}\,)\dot{\varepsilon}^{i}\dot{\varepsilon}^{j}}$, $H$ is the Hamiltonian,
and $\theta^i\equiv\vartheta^i$ are in fact the Noether invariants associated with the symmetry of the Lagrangian 
under the generators of the left action of the group $SU(2)$ on itself (the right-invariant vector fields):
\begin{equation}
Z_{(i)}^{R}=Z_{(i)}^{R\, k}\frac{\partial}{\partial\epsilon^{k}}=\frac{1}{R}\left(\pm |\epsilon_4|\delta_{i}^{k}+
\eta_{\cdot ij}^{k}\epsilon^{j}\right)\,\frac{\partial}{\partial\epsilon^{k}}.\label{simetriaR}
\end{equation}
To be precise, the Lagrangian (\ref{Lagrangiano}) is invariant under the jet-extension of the vector fields (\ref{simetriaR})
acting also on velocities $\dot{\epsilon}^i$ or $\theta^i$ (the subscript refers to its Noether constant):
\be
X_{(\vartheta^i)}=Z^j_{(i)}\frac{\!\!\partial}{\partial\epsilon^j}+
\frac{\partial Z^j_{(i)}}{\partial\epsilon^k}\dot{\epsilon}^k\frac{\!\!\partial}{\partial\dot{\epsilon}^j}
=Z^j_{(i)}\frac{\!\!\partial}{\partial\epsilon^j}+\frac{2}{R}\eta^k_{.in}\theta^n\frac{\!\!\partial}{\partial\theta^k}=
Z(\vec{\varepsilon})^j_{(i)}\frac{\!\!\partial}{\partial\varepsilon^j}+\frac{2}{R}\eta^k_{.in}\vartheta^n\frac{\!\!\partial}{\partial\vartheta^k}
\,.\label{X}
\ee

In the last equality the expression of $X_{(\vartheta^i)}$ has been given in SM. Note that a new variable $\varepsilon_4$ can be introduced in such a way that we also have 
$\vec\varepsilon{\,}^2+\varepsilon_4^2=R^2$, and therefore the solution manifold has the topology of $T^*S^3$. We shall introduce also hyperspherical coordinates (\ref{hypersphericalcoor}), denoted
with the same variables,  and use the same notation $U_\pm$ and $V$ for the local charts of the $S^3$ part of SM.

Unfortunately, the symmetries above, with associated constants of motion $\vartheta^i$, only span half of the solution manifold of our dynamical problem and three more 
Noether invariants (and the corresponding symmetries) are required to fully parametrize the manifold. To complete the parametrization by means of Noether constants, we have to
resort to non-point vector fields (non-jet-extensions) leaving invariant the Poincar\'e-Cartan form associated with the Lagrangian 
(although not the Lagrangian itself, nor even semi-invariant):
\begin{equation}
\Theta_{PC}=\frac{\partial L}{\partial\dot{\epsilon}^{i}}(d\epsilon^{i}-\dot{\epsilon}^{i}dt)+L\, dt=p_{i}d\epsilon^{i}-H\, dt,\label{thetaPC}
\end{equation}
They are much more easily found directly on the solution manifold $SM$, allowing $\Theta_{PC}$ to fall down through the equations of motion. 
In fact, up to a total differential, $\Theta_{PC}$ acquires a simple expression corresponding to a global structure 
(although written in local co-ordinates):
\bea
\Theta_{PC}\approx\Lambda &\equiv& m\vartheta_i\vartheta^{(i)}=\pi_id\varepsilon^i\;\hbox{(Liouville form)}\\ \label{Lambda}
\Omega &\equiv & d\Lambda = md\vartheta_i\wedge\vartheta^{(i)}+\frac{m}{R}\vartheta_i\eta_{\cdot
jk}^{i}\vartheta^{(j)}\wedge\vartheta^{(k)}=d\pi_ i\wedge d\varepsilon^i\;\hbox{(symplectic form)}\,,\label{Omega}
\eea
where $\pi_i$ is defined by $m\vartheta_i\equiv Z^k_{(i)}\pi_k$ or $\pi_i\equiv m\vartheta_i^{(k)}\vartheta_k$.
Now it is straightforward to test that the vector fields
\be
X^{(\varepsilon^i)}=\frac{\!\!\partial}{\partial\pi_i}=\frac{1}{m}Z(\vec{\varepsilon})^i_{(k)}\frac{\!\!\partial}{\partial\vartheta_k}\label{Y}
\ee
are symmetries of the Liouville form (they satisfy $\Omega(X^{(\varepsilon^i)})=d\varepsilon^i$, indeed)  with Noether invariant $\varepsilon^i$.

However, the symmetries (\ref{X}) and (\ref{Y}) do not close a Lie algebra themselves but require a new symmetry, generated by the vector field
\be
X_{(\varepsilon^4)}=\frac{\varepsilon^i}{2R}\frac{\!\!\partial}{\partial\vartheta^i}\,,\label{Z}
\ee
with Noether invariant $\varepsilon^4= \pm |\varepsilon_4|=\pm R\rho(\vec\varepsilon)$. They close the minimal Lie algebra 
generalizing the Heisenberg-Weyl one, 
\begin{eqnarray}
\left\{ \varepsilon^{i},\varepsilon^{j}\right\}  & = & 0\nn\\
\left\{ \varepsilon^{i},\vartheta_{j}\right\}  & = & \frac{1}{mR}\eta_{\cdot jk}^{i}\varepsilon^{k}+\frac{\varepsilon^4}{mR}\delta_{j}^{i}\nn\\
\left\{ \vartheta_{i},\vartheta_{j}\right\}  & = & \frac{2}{mR}\,\eta_{\cdot ij}^{k}\vartheta_{k}\label{PincheAlgebra}\\
\left\{ \varepsilon^{i},\varepsilon^4\right\}  & = & 0\nn\\
\left\{ \vartheta_{i},\varepsilon^4\right\}  & = & \frac{1}{mR}\varepsilon_{i}\,.\nn
\end{eqnarray}

Let us insist in those facts which distinguish the more standard situation of linear
problems, with classical $SM$ governed by the Heisenberg-Weyl structure (for which
canonical quantization is in order), from this essentially non-linear problem: 
(1) Here, the basic symmetries are found only after resorting to the 
Poincar\'e-Cartan form, rather than the Lagrangian, (2) the generators of the
symmetry are in general non-point vector fields, (3) along with the generators
associated with the ``canonical'' variables an extra generator does appear, and (4) 
the topology of the $SM$ is essentially non-trivial (it possesses non-trivial de-Rham 
cohomology).

The present situation is, on the other hand, quite appropriate to tackle the quantization process on the base of
a Group Approach to Quantization \cite{23} (see \cite{Aldaya-review-GAQ} and references therein and very particularly \cite{SU(2)}). 
In this symmetry-based algorithm we start by replacing the classical symplectic manifold, the $SM$, by a Lie 
group $\tilde{G}\equiv\tilde\Sigma(SU(2))$, that is a central extension by $U(1)$ 
of the classical symmetry, and  which results from exponentiating the Lie algebra (\ref{PincheAlgebra}). 
It provides, like any Lie group, two set of canonically defined generators (right- and left-invariant generators), mutually commuting.
One of those represents the group on the space of complex $U(1)$-equivariant function on $\tilde\Sigma(SU(2))$ (that which would be the
prequantization in the sense of Geometric Quantization \cite{Souriau,Kostant,Kirillov}), whereas the other fully reduces the 
representation by means of compatible restrictions, the polarization of the wave functions, thus constituting the true quantization. 
We further explain the process directly throughout the actual calculations.

We proceed to exponentiate (\ref{PincheAlgebra}) (centrally extended by the generator $1$) in a particular way to arrive at the following group law:
\bea
\vec{\varepsilon\,}^{\prime\prime}&=&\frac{1}{R}\left(\pm|\varepsilon_4|\vec{\varepsilon\,}^{\prime}\pm|\varepsilon_4'|\vec{\varepsilon}+
\vec{\varepsilon\,}^{\prime}\wedge\vec{\varepsilon}\right)\nn\\
\vec{\pi}^{\prime\prime}&=&\vec{\pi}^{\prime}+X^{L}(\vec{\varepsilon\,}^{\prime})\vec{\pi}+\frac{1}{R}\vec{\varepsilon\,}^{\prime}\pi_4\nn\\
\pi_4^{\prime\prime}&=&\pi_4^{\prime}+\frac{1}{R}\left(\pm|\varepsilon_4'|\pi_4-\vec{\varepsilon\,}^{\prime}\cdot\vec{\pi}\right)\\ \label{PincheGroup}
\zeta^{\prime\prime}&=&\zeta^{\prime}\zeta
e^{-i\kappa ((\pm\frac{|\varepsilon_4'|}{R}-1)\pi_4-\frac{1}{R}\vec{\varepsilon\,}^{\prime}\cdot\vec{\pi})}\,,\nn
\eea
where 
 $\zeta=e^{i\phi}\in U(1)$ and the $\pm$ signs refer to the chart $U_\pm$ where the corresponding element lies. Here $\kappa\equiv\frac{R}{\hbar}$, with dimensions of inverse of momentum (it can be either positive or negative, both
 signs leading to equivalent groups laws, the sign is a question of convenience and in fact we shall take $\kappa$ to be negative
 later). 
 The parameters $\vec\pi$ and $\pi_4$ have  the dimensions of a momentum, thus a constant $\hbar>0$ with the
 dimensions of an action has  been introduced to keep the exponent dimensionless. 
 

 
From the group law above we recover the classical symmetry (now extended by $U(1)$ accounting for the phase invariance of the 
future wave functions) constituted by the right-invariant vector fields:
\bea Z_{(\varepsilon^i)}^{R}&=&Z_{(i)}^{R\,
k}\frac{\partial}{\partial\varepsilon^{k}}+ \frac{1}{R}\eta_{\cdot
ik}^{j}\pi^{k}\frac{\partial}{\partial\pi^{j}}+\frac{1}{R}\pi_4\frac{\partial}{\partial\pi^{i}}-
\frac{1}{R}\pi_{i}\left(\frac{\partial}{\partial \pi_4}-\kappa\,\Xi\right)\nn\\
Z_{(\pi^i)}^{R}&=&\frac{\partial}{\partial\pi^{i}}\nn\\
Z_{(\pi_4)}^{R}&=&\frac{\partial}{\partial \pi_4} \label{RIVF}\\
Z_{(\zeta)}^{R}&\equiv&\Xi=i(\zeta\frac{\partial}{\partial\zeta}-\bar{\zeta}\frac{\partial}{\partial\bar{\zeta}}),\nn
\eea
These generators close the same algebra (\ref{PincheAlgebra}) except for a trivial redefinition of the Noether invariant $\varepsilon^4$ with a numerical constant.

To recover the generalized Poincar\'e-Cartan form, that is, the quantization $1$-form $\Theta$, we must compute the left-invariant generators
and select the canonical $1$-form, dual of them, in the $U(1)$ direction. That is:
\bea
Z_{(\varepsilon^i)}^{L}&=&Z_{(i)}^{L\, k}\frac{\partial}{\partial\varepsilon^{k}}\nn\\
Z_{(\pi^i)}^{L}&=&Z_{(i)}^{L\, k}\frac{\partial}{\partial\pi^{k}}-
\frac{1}{R}\varepsilon_{i}\left(\frac{\partial}{\partial \pi_4}-\kappa\,\Xi\right)\nn\\
Z_{(\pi_4)}^{L}&=&\frac{1}{R}\left(\pm|\varepsilon_4|\frac{\partial}{\partial \pi_4}+\varepsilon^{i}\frac{\partial}{\partial\pi^{i}}\right)-\kappa(\pm\frac{|\varepsilon_4|}{R}-1)\Xi\\
Z_{(\zeta)}^{L}&=&\Xi=i(\zeta\frac{\partial}{\partial\zeta}-\bar{\zeta}\frac{\partial}{\partial\bar{\zeta}})\,,\nn
\eea
\be
\Theta=-\kappa\left(\frac{\varepsilon_{i}}{R}d\pi^{i}+(\pm\frac{|\varepsilon_4|}{R}-1)d\pi_4\right) +\frac{d\zeta}{i\zeta}.\label{Theta}
\ee
The quantization form generalizes the Poincar\'e-Cartan one in two respects. On the one hand, it is left strictly invariant 
under the quantum symmetry (the centrally-extended version of the classical one) and on the other, it may have a kernel 
(apart from the time evolution which has been disregarded for the time being), to be included in the restrictions to the wave functions (polarization).
In fact, the kernel of (\ref{Theta}) is the characteristic sub-algebra
\be
{\cal G}_{\Theta}=\langle\mathit{Z}_{(\pi_4)}^{L}\rangle,
\ee
so that $d\Theta/Z_{(\pi_4)}^{L}$ reproduces (\ref{Omega}). The Noether invariants are also regained by contracting $\Theta$ with the 
right-invariant vector fields (see \cite{SU(2)}).

The space of complex univariant functions on the entire quantization group, that is, complex functions of the form 
$\varPsi(\zeta,\vec{\varepsilon},\vec{\pi},\pi_4)=\zeta\psi(\vec{\varepsilon},\vec{\pi},\pi_4)$ supports a unitary, though reducible, 
representation of the group $\tilde{G}\equiv\tilde\Sigma(SU(2))$, with respect to the scalar product using the left Haar measure given 
by the exterior product of the left-invariant canonical $1$-form components. To reduce the representation we must impose the 
maximal restriction compatible with the action of the right-invariant generators (to be prompted to quantum operators). 
To this end, a \textit{polarization}, i.e.
a maximal left sub-algebra, containing the characteristic sub-algebra and excluding the central generator, exists:
\begin{equation}
\mathcal{P}_{\rm c}=\langle Z_{(\vec\pi)}^{L},\, Z_{(\pi_4)}^{L}\rangle\,.
\label{pola1}
\end{equation}
Imposing $\mathcal{P}_{\rm c}\varPsi(\zeta,\vec{\varepsilon},\vec{\pi},\pi_4)=0\,$, we can find wave functions on ``configuration space''. 

\section{Momentum space quantization}
\label{Momento}

In \cite{SU(2)} we realized, using the first-order polarization \eqref{pola1}, the quantization of the $S^3$-sigma particle in the ``configuration space representation'' or
``coordinate representation'', since the variables upon which the wave functions depend arbitrarily on are the
parameters $\vec\varepsilon$. In Appendix \ref{ReviewConfiguration} we recall the main results there obtained, and we provide some new technical details needed here\footnote{Note that we have now used a slightly different notation with respect to \cite{SU(2)}.}.

A ``momentum space representation'' can also be achieved from our quantization group by looking for
a polarization subalgebra containing the generators $Z_{\varepsilon^{i}}^{L}$. Imposing these first-order conditions on functions on the group:
\be
Z_{(\vec{\varepsilon})}^{L}\psi(\vec{\varepsilon},\vec{\pi},\pi_4)=0\;\;\Rightarrow \psi\neq\psi(\vec{\varepsilon})\,,
\ee
we obtain a unitary realization of the group on functions $\psi(\vec\pi,\pi_4)$, with an invariant measure \cite{Quasi-invariant-measure}:

\be
d\mu_{\rm m}=i_{X_{\varepsilon^{1}}^{L}}i_{X_{\varepsilon^{2}}^{L}}i_{X_{\varepsilon^{3}}^{L}}\omega=d\pi_1\wedge d\pi_2\wedge d\pi_3\wedge d\pi_4\,.
\ee
Note that this measure is the standard Lebesgue measure in $\mathbb{R}^4$, which is obviously invariant under the Euclidean group $E(4)$, which contains our $SU(2)$-sigma particle group as a subgroup see below). 
However, this representation is reducible and further restrictions have to be imposed to obtain an irreducible representation.

%

Unfortunately, no first-order polarization
subalgebra does exist  containing  
$Z_{(\vec{\varepsilon})}^{L}$ 
and we have to seek in the left-enveloping algebra \cite{HOPolarization1,HOPolarization2}. In fact, we can construct
the following higher-order polarization subalgebra \cite{SU(2)}:
\be
\mathcal{P}^{HO}=\langle Z_{(\varepsilon^i)}^{L},Z_{(\pi_4)}^{LHO}\rangle\,,
\ee
where the higher-order left generator replacing $Z_{(\pi_4)}^L$ is
\be
Z_{(\pi_4)}^{LHO}\equiv \left(Z_{(\pi_4)}^{L}\right)^2 -2i \kappa\,
Z_{(\pi_4)}^L+Z_{(\vec{\pi})}^L\cdot Z_{(\vec{\pi})}^L\,,
\ee
which commutes with $Z_{(\varepsilon^i)}^{L}$ (actually, it commutes with all generators since it turns out to be the Casimir of the Lie algebra).
This extra second-order polarization condition on the $U(1)$-wave functions $\zeta\psi(\vec{\pi},\pi_4)$  acquires a simple form after
using that $Z_{(\pi_4)}^{LHO}$ is a Casimir and does  commute with the entire
algebra and thus can be rewritten as $Z_{(\pi_4)}^{RHO}$, that is,
using the same algebraic expression though in terms of
right-invariant generators. We then obtain:
\be \left[\frac{\partial^2}{\partial\vec{\pi}^2}+\frac{\partial^2}{\partial
\pi_4^2}-2i\kappa \frac{\!\!\partial}{\partial
\pi_4}\right]\psi=0\,.\label{EcHelmholtzRedef}
\ee
With a simple redefinition of the wave function,
\be \psi(\vec{\pi},\pi_4)=e^{i\kappa \pi_4}\phi(\vec{\pi},\pi_4),
\label{redefinicion} \ee
intended to cancel the first-order derivative, we arrive at

%
\be
\left[\frac{\partial^2}{\partial\vec{\pi}^2}+\frac{\partial^2}{\partial
\pi_4^2}+\kappa^2\right]\phi(\vec{\pi},\pi_4)=0  \,,  \label{EcHelmholtz}
\ee
%
%
%
the solutions of which are eigen-functions of the ``Laplacian'' operator in momentum space
$\Delta_\pi=\frac{\partial^2}{\partial\vec{\pi}^2}+\frac{\partial^2}{\partial \pi_4^2}$. 


The right-invariant vector fields (\ref{RIVF}), when restricted to functions $\phi(\vec{\pi},\pi_4)$, provide the quantum operators:
\bea
\hat{\vec k}\phi(\vec{\pi},\pi_4) &\equiv & i\hbar Z^R_{(\vec\varepsilon)} \phi(\vec{\pi},\pi_4) = \frac{i}{\kappa}\left( \vec{\pi}\times \frac{\partial\,}{\partial \vec{\pi}} + \pi_4\frac{\partial\,}{\partial \vec{\pi}} - \vec{\pi}\frac{\partial\,}{\partial \pi_4}   \right)\phi(\vec{\pi},\pi_4)\nn\\
\hat{\vec\varepsilon}\phi(\vec{\pi},\pi_4)&\equiv& i\hbar Z^R_{(\vec\pi)}\phi(\vec{\pi},\pi_4)   = i\hbar \frac{\partial\,}{\partial \vec{\pi}}\phi(\vec{\pi},\pi_4) \label{RealizacionMomentos}\\
\hat{\varepsilon}_4\phi(\vec{\pi},\pi_4)&\equiv&  i\hbar Z^R_{(\pi_4)} \phi(\vec{\pi},\pi_4) = i\hbar \left(\frac{\partial\,}{\partial \pi_4}+i\kappa  \right)\phi(\vec{\pi},\pi_4)\,.\nn
\eea


Those operators realize a representation of the group 
$\tilde\Sigma(SU(2))$ on solutions of \eqref{EcHelmholtz}, to be promoted to a unitary and irreducible representation once a suitable invariant scalar product has been chosen (see below).
This representation can be extended \cite{SU(2)} to a unitary and irreducible representation of the Euclidean group $E(4)$, with the addition of the operators $\hat{\vec J}$:
\begin{equation}
 \hat{\vec J}\phi(\vec{\pi},\pi_4) =  i \hbar \left( \vec{\pi}\times \frac{\partial\,}{\partial \vec{\pi}}\right) \phi(\vec{\pi},\pi_4)
 \label{momangmomentos} \,.
\end{equation}

Note that Eq. (\ref{EcHelmholtz}) is the Helmholtz equation in four dimensions, but realized in momentum space, therefore the usual wave number (with dimension $L^{-1}$) is substituted by $\kappa= \frac{R}{\hbar}$, with dimension inverse of momentum, $M^{-1}L^{-1}T$.

Thus the wave function in momentum representation behaves like an optical wave (in the Helmholtz approximation) in four dimensional space. We can use the full machinery of Helmholtz Optics \cite{BernardoOptica} to describe our wave functions in momentum space, but  assigning a probability interpretation to these wave functions in terms
of a suitable invariant scalar product in a Hilbert space made of solutions of (\ref{EcHelmholtz}).

It should be stressed that Eq. (\ref{EcHelmholtz}) resembles very much the Klein-Gordon equation in four dimensional Minkowski space-time, and this is a consequence of the fact that there is a certain similitude between the Euclidean group
$E(4)$ and the  Poincar\'e group $ISO(3,1)$. There are, however, important differences at the topological (and algebraic) level 
between these two groups and therefore we must be cautious when translating results from the Klein-Gordon equation to the Helmholtz one.

We know from the Klein-Gordon equation that, the metric being  Lorentzian, there is a singularized coordinate ($t$ or $x_0$) and, therefore, it can be seen as a second order differential equation in time (or in $x_0$). To specify a particular solution we need an initial condition for the function itself and for its derivative with respect to time, where the initial condition is usually set at $t=0$ (or $x_0=0$), corresponding to Cauchy boundary conditions. Boundary conditions for the function at two different time values can also be given, corresponding to Dirichlet boundary conditions, although this is in general an ill-posed problem for the Klein-Gordon equation \cite{Payne}. 

For the Helmholtz equation, there is no singularized coordinate (the metric is Euclidean), therefore we can specify Dirichlet or  Newman (or mixed, i.e. Robin) boundary conditions for the solution at the boundary of some four dimensional region (a solid sphere $S^3$, for instance), these being well-posed problems. Another possibility, mimicking the Klein-Gordon case, is to  
arbitrarily singularize a momentum component, $\pi_4$ for instance, and specify Cauchy boundary conditions (i.e. ``initial'' conditions for the function and its derivative at a particular value, like $\pi_4=0$, of this momentum component). Although Cauchy boundary conditions for the Helmholtz equation are ill-posed in general \cite{Payne}, further restrictions can be imposed in order to guarantee
the existence, uniqueness and continuity on the data of the solutions \cite{Steinberg-Wolf}.
%

Even more, the group law given in Eq. (\ref{PincheGroup}) for the extended group singularizes the momentum component $\pi_4$ (as well as the coordinate $\varepsilon_4)$, and this facilitates the use of Cauchy boundary conditions. In essence, we have introduced a virtual hyper-screen (3-dimensional screen) given by $\pi_4=0$, and described the solutions of Helmholtz equation by their values and those of their 
normal derivative at the hyper-screen. The choice of the orientation of this hyper-screen also determines the choice of the local charts $U_\pm$, as those corresponding to the projection to a particular equatorial sphere of $S^3$, namely that determined by $\varepsilon_4=0$.


It should be remarked, however, that the momentum component $\pi_4$ used to specify the Cauchy boundary conditions does not define any dynamics on the system, it is used just to reduce the representation
and to obtain the carrier space of an irreducible (and unitary) representation of the SU(2)-sigma particle group. The true dynamics, and the associated time variable $t$, will be introduced later through the corresponding Hamiltonian.  

\subsection{Vector space of Solutions}

To construct the Hilbert space ${\cal H}_{\rm m}$ defining the representation in momentum space, we shall be restricted to solutions which physically correspond to states describing a particle on the sphere $S^3$, that is,   the vector subspace ${\cal V}_{\rm m}$ of \textit{oscillatory solutions} of the Initial Value Problem (IVP) (see Appendix \ref{AppendixHelmholtz}) :
\be
\left[\frac{\partial^2}{\partial\vec{\pi}^2}+\frac{\partial^2}{\partial
\pi_4^2}+\kappa^2\right]\phi(\vec{\pi},\pi_4)=0\,,\qquad \phi(\vec{\pi},0)=\phicirc(\vec{\pi})\,,\quad \frac{\partial\,}{\partial{\pi_4}}\phi(\vec{\pi},0)=\phibullet(\vec{\pi})\,.
\label{IVP-Helmholtz}
\ee

For convenience, we have used the notation $\phicirc(\vec{\pi})$ and $\phibullet(\vec{\pi})$ for the initial values of $\phi$ and its derivative, respectively, but we have to take into account that 
$\phicirc(\vec{\pi})$ and $\phibullet(\vec{\pi})$ in the IVP (\ref{IVP-Helmholtz}) are (independent) input data.

Note that since the IVP (\ref{IVP-Helmholtz}) is linear, we can decompose ${\cal V}_m$ 
into two linear subspaces: $\Vcirc$, made of oscillatory solutions $\phi$ for which 
$\phibullet=0$,  and $\Vbullet$,  made of oscillatory solutions $\phi$ for which $\phicirc=0$.
Any oscillatory solution $\phi$ of the IVP (\ref{IVP-Helmholtz}) can be uniquely decomposed as $\phi=\phi^{(\circ)}+\phi^{(\bullet)}$, with $\phi^{(\circ)}\in \Vcirc$ and $\phi^{(\bullet)}\in \Vbullet$. Let us define the projectors $\Pcirc$ and $\Pbullet$ associated with this decomposition. These projectors
are idempotent and satisfy $\Pcirc \Pbullet=\Pbullet\Pcirc=0$ and $\Pcirc + \Pbullet=I_{{\cal V}_{\rm m}}$. Therefore, we can decompose ${\cal V}_{\rm m}$ as the direct sum:
\be
{\cal V}_{\rm m}= \Vcirc \oplus \Vbullet\,.
\label{sumaV}
\ee
At this level, the direct sum is one of vector spaces. Later, we shall introduce an invariant scalar product, and we shall see that the projectors
$\Pcirc$ and $\Pbullet$ are self-adjoint with respect to this scalar product and therefore the direct sum appearing in eq. (\ref{sumaV}) will be also one of orthogonal subspaces.

\subsection{Construction of an invariant scalar product}

Let us construct an invariant scalar product for functions in momentum space. For that purpose, we shall follow\footnote{An alternative construction is possible by using standard Fourier Analysis in
$\mathbb{R}^4$ \cite{Gustavo}.} \cite{Steinberg-Wolf} and introduce the most general sesquilinear form involving both
the functions and their derivatives with respect to $\pi_4$ (we shall continue to singularize $\pi_4$ in our IVP):

\be
\langle \phi,\phi'\rangle_{\rm m}=\int_{\mathbb{R}^3}d\vec{\pi}\int_{\mathbb{R}^3}d\vec{\pi}{\,}' (\phi(\vec\pi,\pi_4)^*,\frac{\partial\,}{\partial \pi_4} \phi(\vec\pi,\pi_4)^*)
\left(\begin{array}{cc}
\stackrel{\circ\circ}{K}(\vec\pi,\vec\pi{\,}') & \stackrel{\circ\bullet}{K}(\vec\pi,\vec\pi{\,}') \\
\stackrel{\bullet\circ}{K}(\vec\pi,\vec\pi{\,}') & \stackrel{\bullet\bullet}{K}(\vec\pi,\vec\pi{\,}')
\end{array}\right)
\left(
\begin{array}{c}
\phi'(\vec\pi{\,}',\pi_4')\\ \frac{\partial\,}{\partial \pi_4'} \phi'(\vec\pi{\,}',\pi_4')
\end{array} \right)\,.
\ee

Invariance under the generator $Z^R_{(\vec\pi)}$ in (\ref{RealizacionMomentos}) imposes  $\stackrel{\cdot\cdot}{K}\!\!(\vec\pi,\vec\pi{\,}')=\stackrel{\cdot\cdot}{K}\!\!(\|\vec\pi-\vec\pi{\,}'\|)$, where $\cdot=\circ,\bullet$.
Imposing invariance under the generators $Z^R_{(\pi_4)}$ and $Z^R_{(\vec\varepsilon)}$  in (\ref{RealizacionMomentos}) leads to $\stackrel{\circ\bullet}{K}=\stackrel{\bullet\circ}{K}=0$ and 
\be
\stackrel{\circ\circ}{K}(\|\vec\pi-\vec\pi{\,}'\|)= \frac{J_2(\kappa \|\vec\pi-\vec\pi{\,}'\|)}{\|\vec\pi-\vec\pi{\,}'\|^2}\equiv\kappa^2 k_2(\kappa(\vec\pi-\vec\pi{\,}'))\,,\qquad 
\stackrel{\bullet\bullet}{K}(\|\vec\pi-\vec\pi{\,}'\|)=\frac{J_1(\kappa \|\vec\pi-\vec\pi{\,}'\|)}{\kappa\|\vec\pi-\vec\pi{\,}'\|} \equiv k_1(\kappa(\vec\pi-\vec\pi{\,}'))\,,
\ee
where $J_\alpha(z)$ are the Bessel functions of the first kind and the functions $k_\alpha$ are defined in Appendix \ref{AppendixBessel}. Thus, up to a constant $C>0$, we can write
\begin{equation}
	\langle \phi,\phi'\rangle_{\rm m} = 
	C\int_{\mathbb{R}^3}d\vec{\pi}\int_{\mathbb{R}^3}d\vec{\pi}{\,}' \left( \phi(\vec\pi,\pi_4)^*  \kappa^2 k_2(\kappa \|\vec\pi-\vec\pi{\,}'\|) \phi{}'(\vec\pi{\,}',\pi_4')  +\frac{\partial\,}{\partial \pi_4}\phi(\vec\pi,\pi_4)^*  k_1(\kappa\|\vec\pi-\vec\pi{\,}'\|) \frac{\partial\,}{\partial \pi_4'}\phi{}'(\vec\pi{\,}',\pi_4') \right)\,. 
	\label{ProductoEscalarMomentosarriba}
\end{equation}

%
 
Due to the invariance under  $Z^R_{(\pi_4)}$, the scalar product $\langle\cdot,\cdot\rangle_{\rm m}$ is invariant under the ``$\pi_4$ evolution'', and thus it can be written as:
\bea
\langle \phi,\phi'\rangle_{\rm m}&=& C\int_{\mathbb{R}^3}d\vec{\pi}\int_{\mathbb{R}^3}d\vec{\pi}{\,}' \left( \phicirc(\vec\pi)^*   \kappa^2 k_2(\kappa \|\vec\pi-\vec\pi{\,}'\|) \phicirc{}'(\vec\pi{\,}') 
+ \phibullet(\vec\pi)^* k_1(\kappa \|\vec\pi-\vec\pi{\,}'\|) \phibullet{}'(\vec\pi{\,}') \right) \nonumber \\
& & \equiv \kappa^2\langle \phicirc,\phicirc{}'\rangle_{\hat{K}_2}+ \langle \phibullet,\phibullet{}'\rangle_{\hat{K}_1}
\,,\label{ProductoEscalarMomentos}
\eea
expressing the scalar product in terms of just the input data of (\ref{IVP-Helmholtz}), and showing that it can be written as the sum of two scalar products of the type introduced in Appendix \ref{DiracNotation}.

Since we have restricted to \textit{oscillatory solutions}  of \eqref{IVP-Helmholtz},  the operators $\hat{K}_2$ and $\hat{K}_1$ are positive definite (see Appendix \ref{AppendixBessel}) and therefore
the scalar product is positive definite. Note that these \textit{oscillatory solutions} are precisely those whose configuration space counterpart 
 are well-defined functions on the sphere $S^3$ (see Sec. \ref{Fourier}, Appendix \ref{AppendixHelmholtz} and \cite{Steinberg-Wolf}). 

Restricting to the subspace of normalizable (with respect to $\langle\cdot,\cdot\rangle_{\rm m}$)  solutions in ${\cal V}_{\rm m}$, and completing this subspace with respect to 
the scalar product $\langle\cdot,\cdot\rangle_{\rm m}$, we obtain the Hilbert space ${\cal H}_{\rm m}$, which is the carrier space of a unitary and irreducible representation of the group $\tilde\Sigma(SU(2))$\footnote{This representation can be extended to a unitary and irreducible representation of the whole Euclidean group $E(4)$, see  \eqref{momangmomentos}.}.

Is is worth mentioning that this scalar product involves a double integral in $\vec\pi$ and $\vec\pi{\,}'$ (i.e. it is a ``non-local'' scalar product;  it is not a standard Lebesgue integral), and contains non-trivial kernels $\frac{J_2(\kappa \|\vec\pi-\vec\pi{\,}'\|)}{\|\vec\pi-\vec\pi{\,}'\|^2}$ and 
$\frac{J_1(\kappa \|\vec\pi-\vec\pi{\,}'\|)}{\kappa\|\vec\pi-\vec\pi{\,}'\|}$ which depend on $\|\vec\pi-\vec\pi{\,}'\|$, so that they are convolution kernels implying that the scalar product can be ``diagonalized'' (i.e. transformed
into a ``local'', or Lebesgue, scalar product) by a Fourier transform. We shall discuss this point in Sec. \ref{Fourier} (see also Appendices \ref{DiracNotation} and \ref{AppendixBessel} for further details). 
f

From the form of the scalar product (\ref{ProductoEscalarMomentos}), it is clear that the subspaces $\Vcirc$ and $\Vbullet$ are orthogonal to each other. Defining $\Hcirc={\cal H}_{\rm m}\cap \Vcirc$ and
$\Hbullet={\cal H}_{\rm m}\cap \Vbullet$, we have that  ${\cal H}_{\rm m}=\Hcirc\oplus \Hbullet$, where now this is a direct sum of (orthogonal) Hilbert subspaces.

Note that we can identify $\Hcirc$ and $\Hbullet$ with corresponding subspaces 
$\overline{\Hc^{(e)}}$ and $\Hc^{(o)}$ in configuration space, respectively (see Appendix \ref{ReviewConfiguration} and Sec. \ref{Fourier}).
That is,  $\Hcirc$ corresponds to even functions  on the sphere $S^3$ with respect to the equatorial sphere $S^2$ at $\varepsilon_4=0$ and $\Hbullet$ corresponds to odd functions on the sphere $S^3$.

\subsection{Basic operators and Hermiticity} 

Given the Hilbert space structure of momentum space representation, we redefine 
the vector fields \eqref{RealizacionMomentos} in order to obtain quantum 
operators, Hermitian with respect to scalar product $\langle\cdot,\cdot\rangle_m$ 
in \eqref{ProductoEscalarMomentosarriba}:

\bea
\hat{\vec k}\phi(\vec{\pi},\pi_4)&=&  i\hbar Z^R_{(\vec\varepsilon)} \phi(\vec{\pi},\pi_4) =    \frac{i}{\kappa}\left( \vec{\pi}\times \frac{\partial\,}{\partial \vec{\pi}} + \pi_4\frac{\partial\,}{\partial \vec{\pi}} - \vec{\pi}\frac{\partial\,}{\partial \pi_4}  \right)\phi(\vec{\pi},\pi_4)\nn\\
\hat{\vec\varepsilon}\phi(\vec{\pi},\pi_4)&=&  i\hbar Z^R_{(\vec\pi)}\phi(\vec{\pi},\pi_4)   =  i\hbar  \frac{\partial\,}{\partial \vec{\pi}} \phi(\vec{\pi},\pi_4) \label{PincheRedmomentos}\\
 \hat{\varepsilon}_{4}\phi(\vec{\pi},\pi_4)&=&     \left(i\hbar Z^R_{(\pi_4)}+R \right) \phi(\vec{\pi},\pi_4) =  i\hbar  \frac{\partial\,}{\partial \pi_4}\phi(\vec{\pi},\pi_4).\nn  
\eea
Physically, they represent  the three-component generator of translations on the sphere $S^3$ ($\hat{\vec k}$, also angular momentum generating rotations in ambient space changing the north pole) and position on the sphere in momentum space 
 ($\hat{\vec\varepsilon}$ and $\hat{\varepsilon}_{4}$), whereas $\hat{\vec J}$ in 
 \eqref{momangmomentos} represents angular momentum on the sphere $S^3$ (also generates rotations around the north pole). 
 The interpretation of Helmholtz equation \eqref{IVP-Helmholtz} 
is now transparent: it encodes, in momentum space, the restriction of ``being'' on a
sphere $S^3$ with radius $R=\hbar \kappa$, provided by the operator equation
\[
\hat{\vec\varepsilon\,}^2 + \hat{\varepsilon}_{4}^2 = R^2\,.
\]
Note that the requirement in configuration space of real coordinates $\vec\varepsilon$, 
$\varepsilon_4$ is, in momentum space, that of $\hat{\vec\varepsilon}$ and 
$\hat{\varepsilon}_{4}$ being Hermitian on physical solutions. This amounts to restricting to \textit{oscillatory} solutions of the Helmholtz equation.  

It is instructive to check Hermiticity for those operators explicitly. For 
$\hat{\vec\varepsilon}$, it follows immediately by integrating by parts. To show 
Hermiticity for $\hat{\varepsilon}_{4}$, recurrence relations for the integral kernel 
of the scalar product are needed (namely, 
$2 \frac{d J_\alpha(x)}{dx} = J_{\alpha-1}(x)-J_{\alpha+1}(x)$ and 
$2 \alpha \frac{J_\alpha(x)}{x} = J_{\alpha-1}(x)-J_{\alpha+1}(x)$). Let us compute 
(we drop wave function arguments)

\begin{align*}
	&\langle \phi,\hat{\varepsilon}_{4}\phi'\rangle_m =
	 i\hbar C \int_{\mathbb{R}^3}d{\vec\pi}\int_{\mathbb{R}^3}d{\vec\pi}{}' \left( \phi^*  \kappa^2 k_2(\kappa \|\vec\pi-\vec\pi'\|) \frac{\partial \phi{}'}{\partial \pi_4'} 
 - \frac{\partial \phi^*}{\partial \pi_4}  k_1(\kappa\|\vec\pi-\vec\pi'\|) (\frac{\partial^2 \phi{}'}{\partial \vec{\pi}'^2}+\kappa^2\phi{}') \right)
 \\
 &= i\hbar C \int_{\mathbb{R}^3}d{\vec \pi}\int_{\mathbb{R}^3}d{\vec\pi}{}' \left( \phi^*  \kappa^2 k_2(\kappa \|\vec\pi-\vec\pi'\|) \frac{\partial \phi{}'}{\partial \pi_4'} 
 -  \kappa^2\frac{\partial \phi^*}{\partial \pi_4}  k_1(\kappa\|\vec\pi-\vec\pi'\|) \phi{}'\right.
 - \left.\frac{\partial \phi^*}{\partial \pi_4} 
 \kappa^2 k_2(\kappa\|\vec\pi-\vec\pi'\|)
 (\pi_i-\pi'_i) \frac{\partial \phi{}'}{\partial \pi_i'}\right)
 \\
 &= i\hbar C \int_{\mathbb{R}^3}d{\vec\pi}\int_{\mathbb{R}^3}d{\vec\pi}{}' \left( \kappa^2 k_2(\kappa \|\vec\pi-\vec\pi'\|)\Big(\phi^* \frac{\partial \phi{}'}{\partial \pi_4'} - \frac{\partial \phi^*}{\partial \pi_4}\phi{}'\Big)
\right)
 \,, 
\end{align*}
where we have used the Helmholtz equation in the first equality,  in the second one we 
have integrated by parts in $\pi'_i$ and used recurrence relations, and in the third 
one we have again integrated by parts in $\pi'_i$ and used recurrence relations. In a 
similar way, computing $\langle \hat{\varepsilon}_{4}\phi,\phi'\rangle_m$ leads to 
the same result (integrations by parts are now made in $\pi_i$), showing Hermiticity of $\hat{\varepsilon}_{4}$. Hermiticity for $\vec{\hat k}$ (and $\hat{\vec J}$) also follows by using the same relations. However, it must be pointed out that Hermiticity is guaranteed given that operators $\hat K_1$ and $\hat K_2$ of the integral kernel in the scalar product have been chosen to commute with basic operators \eqref{RealizacionMomentos} in order to achieve invariance (see equation \eqref{kadjunto} from Appendix \ref{DiracNotation}).
Physically this means that the scalar product, being  also invariant under finite transformations, is invariant under translations and rotations of the hyperplane where initial data are given ($\pi_4=0$ in our case).

\subsection{Basis of position eigenstates  in momentum space representation}

Let us introduce in momentum space  a basis of eigenstates of the  commuting operators $\hat{\vec\varepsilon}$ and $\hat{\varepsilon}_{4}$ in \eqref{PincheRedmomentos}. The wave functions of these eigenstates 
must also  satisfy  Helmholtz equation (\ref{EcHelmholtz}).


The spectrum of $\hat{\vec\varepsilon}$  is continuous and doubly degenerated, so that the associated eigenfunctions will be distributions:
\begin{equation}
\phi_{\vec{\varepsilon}\pm}(\vec{\pi},\pi_4)= e^{-\frac{i}{\hbar}\left(\vec{\varepsilon}\cdot \vec{\pi}\pm |\varepsilon_4|\pi_4\right)}\,,
\label{elemsols}
\end{equation}
where $|\varepsilon_4|=\sqrt{R^2-\vec{\varepsilon}{\,}^2}$, as before. We have to impose $\|\vec\varepsilon\|\leq R$ in order to have ``oscillatory'' solutions, for which the IVP  (\ref{IVP-Helmholtz}) is well-posed
and the scalar product is well-defined (see above and Appendix \ref{AppendixHelmholtz}). Note also that these eigenstates  are the momentum space version of the
localized states $\phi_{\vec{\varepsilon}\pm}$ (or $|\vec{\varepsilon}\pm\rangle$) on the sphere $S^3$ (see \eqref{PositionEigenstatesConf} in Appendix \ref{ReviewConfiguration}; see below for the relation between these two sets of functions through the generalized Fourier transform).

These solutions can be decomposed according to (\ref{sumaV}) as:
\begin{equation}
 \phi^{(\circ)}_{\vec{\varepsilon}\pm}=\frac{1}{2}\left(\phi_{\vec{\varepsilon}+}+\phi_{\vec{\varepsilon}-}\right)= e^{-\frac{i}{\hbar}\vec{\varepsilon}\cdot \vec{\pi}}\cos (\frac{1}{\hbar}|\varepsilon_4|\pi_4) \,,\qquad 
 \phi^{(\bullet)}_{\vec{\varepsilon}\pm}=\pm\frac{1}{2}\left(\phi_{\vec{\varepsilon}+}-\phi_{\vec{\varepsilon}-}\right) =\mp i \,e^{-\frac{i}{\hbar}\vec{\varepsilon}\cdot \vec{\pi}} \sin (\frac{1}{\hbar}|\varepsilon_4|\pi_4)\,.
\end{equation}
Note that $\phi^{(\circ)}_{\vec{\varepsilon}-}= \phi^{(\circ)}_{\vec{\varepsilon}+}\equiv \phi^{(\circ)}_{\vec{\varepsilon}}$ and 
$\phi^{(\bullet)}_{\vec{\varepsilon}-}=-\phi^{(\bullet)}_{\vec{\varepsilon}+}\equiv- \phi^{(\bullet)}_{\vec{\varepsilon}}$\footnote{These states $\phi^{(\circ)}_{\vec{\varepsilon}}$ and $\phi^{(\bullet)}_{\vec{\varepsilon}}$ are the momentum space version of even and odd states (Schr\"odinger's cat states) introduced in configuration space, respectively (see (\ref{evenoddconf})).}.

\subsection{Comparison with the Poincar\'e group}
\label{ComparisonPoincare}

Let us remark the main differences with the Klein-Gordon case \cite{Steinberg-Wolf}. For the Klein-Gordon case $\stackrel{\circ\circ}{K}=\stackrel{\bullet\bullet}{K}=0$ and $\stackrel{\circ\bullet}{K}=-\stackrel{\bullet\circ}{K}=\frac{i}{2} \delta(\vec x-\vec x{\,}')$,  resulting in the scalar product:
\bea
\langle \phi,\phi'\rangle^{KG} &=&\frac{i}{2} \int_{\mathbb{R}^3}d\vec x \left( \phi(\vec x,x_0)^*  \frac{\partial\,}{\partial x_0}\phi{}'(\vec x,x_0) - 
\phi'(\vec x,x_0)  \frac{\partial\,}{\partial x_0}\phi{}(\vec x,x_0)^* \right) \nonumber \\  
&=& \frac{i}{2} \int_{\mathbb{R}^3}d\vec x \left( \phicirc(\vec x)^* \phibullet{}'(\vec x) - \phibullet(\vec x)^* \phicirc{}'(\vec x) \right) \,,
\eea
due to invariance under time evolution. This scalar product is positive definite if  it is restricted to the invariant subspace of positive energy solutions of the Klein-Gordon equation, and this constitutes the carrier space of a unitary and irreducible representation
of the orthochronous Poincar\'e group. It is a ``local'' scalar product (i.e. involves a single integral in $\vec x$, a consequence of the fact that the non-zero kernels are Dirac deltas), and it is skew-symmetric in
$\phicirc$ and $\phibullet$. These are important differences with dramatic consequences. However, there is another important difference due to the fact that for positive energy solutions
we can take the ``square root'' of Klein-Gordon equation to obtain a Schr\"odinger-like equation \cite{PositionOperator}, known as the free Salpeter equation \cite{Salpeter,Salpeter2,KowalskiSalpeter}:
\be
i\hbar \frac{\partial\,}{\partial x_0}\phi(\vec x,x_0) = \hat{P}_0 \phi(\vec x,x_0) \equiv \sqrt{m^2c^2-\hbar^2\vec\nabla_{\vec x}^2} \,\,\phi(\vec x,x_0) \,,
\ee
and this implies that $i\hbar \!\!\phibullet=\hat{P}_0 \!\phicirc$ (i.e. the two initial conditions are not independent). 

\subsection{Paraxial Approximation}

If in equation (\ref{EcHelmholtzRedef}) we assume $|\frac{\partial^2\psi}{\partial\pi_4^2}|<<|\kappa \frac{\partial\psi}{\partial\pi_4}|$, Helmholtz equation becomes a Schr\"odinger-like equation:
\begin{equation}
 -i\kappa \frac{\partial\psi}{\partial \pi_4} =-\frac{1}{2} \frac{\partial^2\psi}{\partial\vec{\pi}^2}\,,\label{ParaxialApprox}
\end{equation}
with $\vec\pi$ playing the role of position and $-\kappa\pi_4$ that of time. This approximation is similar to that of the non-relativistic limit taking the Klein-Gordon equation into the Schr\"odinger equation. 

The analogue of this in Helmholtz Optics  corresponds to light rays propagating almost parallel to the optical axis  in the forward direction, known as  Paraxial Approximation.
Although this subspace of solutions is not invariant under the Euclidean group, it has a physical meaning and is a common construction at the laboratory.  For these subspaces the expectation value of $\hat{\varepsilon}_{4}$ is both positive and has a gap, and therefore
a construction similar to the Klein-Gordon case is possible \cite{BernardoManko}. Note however, that there are two different paraxial approximations, corresponding to forward and backward propagation along the optical axis, the backward propagation corresponding to taking 
a negative value of $\kappa$, and therefore there are two limiting Schr\"odinger-like equations. In fact, it can be shown \cite{BernardoManko} that ${\cal H}_{\rm m}\approx L^2_+(\mathbb{R}^3)\oplus L^2_-(\mathbb{R}^3)$,
where $L^2_\pm(\mathbb{R}^3)$ are the carrier spaces of the forward/backward Schr\"odinger equation.

\section{Generalized Fourier Transform}
\label{Fourier}

In this Section we show, by constructing a generalized Fourier transform between 
configuration and momentum representations, that both representation spaces are 
actually unitarily equivalent, provided we restrict the momentum representation to 
\textit{oscillatory} solutions of the Helmholtz equation, as already pointed out. This 
condition is fulfilled when any wave function in momentum space can be expanded in 
terms of the elementary solutions \eqref{elemsols}:
\begin{equation}
\phi_{\vec{\varepsilon},\pm}(\vec{\pi},\pi_4)= N e^{-\frac{i}{\hbar}
\left(\vec{\varepsilon}\cdot \vec{\pi}\pm |\varepsilon_4|\pi_4\right)}\,,
\end{equation} 
where $|\varepsilon_4|=\sqrt{R^2-\vec{\varepsilon}{\,}^2}$,
$\vec\varepsilon\in B_R$, and we have now made a normalizing constant explicit, to be taken $N=\frac{1}{\sqrt{2}(2\pi\hbar)^{3/2}} $ for later convenience. These 
are simultaneous eigenstates of $\hat{\vec \varepsilon}$ and $\hat\varepsilon_{4}$ 
and physically represent states of well-defined position on $S^3$. We can check their 
overlapping by making use of the scalar product \eqref{ProductoEscalarMomentos}, in 
which we now set $C=\frac{\hbar \kappa^2}{4 \pi}$.
Note that input data for solving Helmholtz equation, associated with 
$\phi_{\vec{\varepsilon},\pm}(\vec{\pi},\pi_4)$, are given by
\begin{equation}
	\phicirc_{\vec{\varepsilon},\pm}(\vec \pi) = 
	N e^{-\frac{i}{\hbar} \vec{\varepsilon}\cdot \vec{\pi}}\,,
	\qquad
	\phibullet_{\vec{\varepsilon},\pm}(\vec \pi) = 
	\mp  \frac{i}{\hbar} N|\varepsilon_4|
	e^{-\frac{i}{\hbar} \vec{\varepsilon}\cdot \vec{\pi}}\,.
	\label{inputposicion}
\end{equation}
With that:
\begin{align*}
&\langle \phi_{\vec{\varepsilon},\sigma} , \phi_{\vec{\varepsilon}\,',\sigma'} \rangle_m 
=
\\ 
& C|N|^2 \int_{\mathbb{R}^3} d\vec\pi \int_{\mathbb{R}^3} d\vec\pi' 
\left(\kappa^2 k_2(\kappa \|\vec\pi-\vec\pi'\|) 
e^{\frac{i}{\hbar}\vec{\varepsilon}\cdot \vec{\pi}}
e^{-\frac{i}{\hbar} \vec{\varepsilon}\,'\cdot \vec{\pi}\,'}
+  \frac{\sigma \sigma'}{\hbar^2}|\varepsilon_4||\varepsilon_4'|
k_1(\kappa \|\vec\pi-\vec\pi'\|)  
e^{\frac{i}{\hbar}\vec{\varepsilon}\cdot \vec{\pi}}
e^{-\frac{i}{\hbar} \vec{\varepsilon}\,'\cdot \vec{\pi}\,'}
\right)=
\\
&C|N|^2 \int_{\mathbb{R}^3} d\vec\pi' 
\left\{
\int_{\mathbb{R}^3} d\vec\pi 
\left(\kappa^2 k_2(\kappa \|\vec\pi-\vec\pi'\|) 
+ \frac{\sigma \sigma'}{\hbar^2}|\varepsilon_4||\varepsilon_4'|
k_1(\kappa \|\vec\pi-\vec\pi'\|) 
\right)
e^{\frac{i}{\hbar} \vec{\varepsilon}\cdot (\vec{\pi}-\vec{\pi}\,')}
\right\}
e^{-\frac{i}{\hbar} (\vec{\varepsilon}\,'-\vec{\varepsilon})\cdot \vec{\pi}\,'}=
\\
& \!\!\!\!\!\!\!\!\!\!\!\!\!\!\!\!\frac{4 \pi}{\hbar \kappa^2}C|N|^2
 \int_{\mathbb{R}^3} d\vec \pi' 
\left(
(|\varepsilon_4|)_+ +\sigma \sigma' |\varepsilon_4||\varepsilon_4'|(|\varepsilon_4|)^{-1}_+ 
\right)
e^{-\frac{i}{\hbar} (\vec{\varepsilon}\,'-\vec{\varepsilon})\cdot \vec{\pi}\,'}
\!\!\!=
\!\frac{1}{2} (1	+\sigma \sigma') (|\varepsilon_4|)_+\delta(\vec \varepsilon - \vec \varepsilon\,') =
\delta_{\sigma \sigma'}(|\varepsilon_4|)_+\delta(\vec \varepsilon - \vec \varepsilon\,')
\,, 
\end{align*}
where $\sigma,\sigma'=+,-$, $\int_{\mathbb{R}^3}d\vec \pi 
e^{-\frac{i}{\hbar} \vec \varepsilon \cdot \vec \pi } = 
(2\pi \hbar)^3\delta(\vec\varepsilon)$ 
and, from Appendix \ref{AppendixBessel},  
$\int_{\mathbb{R}^3} d\vec \pi 
k_2(\kappa\|\vec \pi\|)
e^{-\frac{i}{\hbar} \vec \varepsilon \cdot \vec \pi } =  
\frac{4 \pi}{\hbar\kappa^4} (|\varepsilon_4|)_+$, 
$ \int_{\mathbb{R}^3} d\vec \pi k_1(\kappa\|\vec \pi\|)
e^{-\frac{i}{\hbar} \vec \varepsilon \cdot \vec \pi } = 
\frac{4 \pi \hbar(|\varepsilon_4|)^{-1}_+}{\kappa^2}$, where 
$(|\varepsilon_4|)_+$ and $(|\varepsilon_4|)^{-1}_+$ are zero if $\vec\varepsilon \notin B_R$ (see Appendix \ref{AppendixBessel}).  We see that equation \eqref{overlapposicion} in Appendix \ref{ReviewConfiguration} is reproduced.

So, an arbitrary state $\psi\in {\cal H}_m$ can be written:
\begin{equation}
	\psi(\vec \pi,\pi_4)= \int_{B_R}\frac{d\vec\varepsilon}{|\varepsilon_4|}
\left(\Psi^+(\vec\varepsilon\,)  \phi_{\vec{\varepsilon},+}(\vec \pi,\pi_4)  + 
      \Psi^-(\vec\varepsilon \,)  \phi_{\vec{\varepsilon},-}(\vec \pi,\pi_4)\right)\,, \label{FourierUp}
\end{equation}
where the coefficients of the expansion are
$\Psi^\pm(\vec\varepsilon \,)=\langle \phi_{\vec{\varepsilon},\pm},\psi\rangle_m$, 
that is, they can be interpreted as the corresponding wave functions components in 
configuration space, say $\psi^\pm (\vec\varepsilon \,)=\Psi^\pm(\vec\varepsilon \,)$ 
(see \eqref{ResolutionIdentityConf2}). 
 
 For a $\psi\in {\cal H}_m$ written as above, input data are given by: 
 \begin{align}
 	\psicirc (\vec\pi) &= 
 	\frac{1}{\sqrt{2}(2\pi\hbar)^{3/2}} \int_{B_R}\frac{d\vec\varepsilon}{|\varepsilon_4|}
\left(\psi^+(\vec\varepsilon\,) + \psi^-(\vec\varepsilon\,)\right)
e^{-\frac{i}{\hbar} \vec{\varepsilon}\cdot \vec{\pi}} \,,
\label{fourier1}
\\
\psibullet(\vec\pi) &= 
 \frac{-i}{\sqrt{2}\hbar(2\pi\hbar)^{3/2}} \int_{B_R}d\vec\varepsilon
\left(\psi^+(\vec\varepsilon\,) - \psi^-(\vec\varepsilon\,)\right)
e^{-\frac{i}{\hbar} \vec{\varepsilon}\cdot \vec{\pi}} \,.
\label{fourier2}
\end{align}
Those expressions constitute a generalized Fourier transform from configuration space 
state components to momentum space state components. It is instructive, for further 
convenience, to express the previous equations in terms of the (standard 3D) Fourier 
components $(\hat{\psicirc},\hat{\psibullet})$ of $(\psicirc,\psibullet)$:
 \begin{align}
 	\hat\psicirc (\vec\varepsilon) &= 
 	\frac{1}{\sqrt{2}} (|\varepsilon_4|)_+^{-1}
\left(\psi^+(\vec\varepsilon\,) + \psi^-(\vec\varepsilon\,)\right)
\,,
\label{fourier3}
\\
\hat\psibullet(\vec\varepsilon) &= \frac{-i}{\sqrt{2}\hbar} (|\varepsilon_4|)_+^{0}
\left(\psi^+(\vec\varepsilon\,) - \psi^-(\vec\varepsilon\,)\right)\,.
\label{fourier4}
\end{align}
where $(|\varepsilon_4|)_+^{0}\equiv\chi_{B_R}(\vec\varepsilon)$ is the indicator function of $B_R$, and the functions $(|\varepsilon_4|)_+^{-1}$ and $(|\varepsilon_4|)_+^{0}$ appear since the integration in $\vec\varepsilon\,$ in Eqs. \eqref{fourier1}-\eqref{fourier2} involves only $B_R$, not the whole $\mathbb{R}^3$, i.e. 
\begin{equation}
\int_{B_R}d\vec\varepsilon{\,}' f(\vec\varepsilon{\,}')\delta(\vec\varepsilon-\vec\varepsilon{\,}')= f(\vec\varepsilon)\chi_{B_R}(\vec\varepsilon)\,.
\end{equation}

In order to find the corresponding inverse transformation, we can simply invert the previous relations:
\begin{equation}
\label{inverseFTFourier}
	\psi^\pm(\vec \varepsilon)  =\frac{1}{\sqrt{2}} \left( (|\varepsilon_4|)_+\hat{\psicirc}(\vec \varepsilon)\pm i\hbar (|\varepsilon_4|)_+^0
\hat{\psibullet}(\vec \varepsilon)\right)
 \,. 
\end{equation}

However, the interpretation of this expression is  a bit obscure. 
In order to gain more insight, we shall try to 
parallel the computations above: we would
need an expansion of the wave function in configuration space of an 
arbitrary state in terms of eigenstates of a certain
\textit{momentum operator}. However, such an operator is not present in the 
algebra of basic operators corresponding to $\tilde\Sigma(SU(2))$. We 
proceed arguing that, even though we do not have eigenstates of a \textit{momentum 
operator}, we might be able to find states playing an analogous role. For
that, we realize that \eqref{inputposicion} appear in the twofold 
transformation \eqref{fourier1} and \eqref{fourier2} as the integral 
kernels, and analyze which states have those expressions 
(complex-conjugated) as wave functions in configuration space\footnote{From 
now on, we shall make extensive use of results in Appendix
 \ref{ReviewConfiguration} for configuration space.}. Let us define: 
\begin{equation}
\phi^{\pm}_{\vec \pi \circ}(\vec \varepsilon) \equiv M  
\;e^{\frac{i}{\hbar} \vec \pi \cdot \vec \varepsilon } \,,
\qquad	
\phi^{\pm}_{\vec \pi \bullet}(\vec \varepsilon) \equiv \pm  \frac{i}{\hbar} M' 
|\varepsilon_4| e^{\frac{i}{\hbar} \vec{\pi}\cdot \vec{\varepsilon}}\,,
\label{estadosnuconfig}
\end{equation}
where the $\pm$ sign indicates the component of the wave function in 
configuration space (that is, 
$\phi_{\vec \pi \circ}(\vec \varepsilon)= 
(\phi^{+}_{\vec \pi \circ}(\vec \varepsilon)
,\phi^{-}_{\vec \pi \circ}(\vec \varepsilon))$ and equivalently for 
$\phi_{\vec \pi \bullet}(\vec \varepsilon)$);  
$\vec \pi \circ$, $\vec \pi \bullet$ label the state; $M$ and $M'$ are 
normalizing constants.  It is important to note 
that  ``plane waves'' $\phi_{\vec \pi \circ}(\vec \varepsilon)$ alone are not 
enough to expand the whole Hilbert space $\mathcal{H}_c$: they are even
functions  with respect to reflection on the equatorial sphere $S^2$, i.e.,
they might span $\mathcal{H}_{c(e)}$ at much. On the other hand, 
$\phi_{\vec \pi \bullet}(\vec \varepsilon) \in \mathcal{H}_{c(o)}$, that is, they 
are odd functions with respect to the equator. 

We now compute the overlapping between those $\vec \pi$ states in configuration 
space: 

\begin{align*}
\langle \phi_{\vec \pi \circ} , \phi_{\vec \pi \,' \circ} \rangle_c 
&=
|M|^2\int_{B_R} \frac{d\vec \varepsilon}{|\varepsilon_4|}
\Big(
e^{-\frac{i}{\hbar} \vec \pi \cdot \vec \varepsilon}
e^{\frac{i}{\hbar} \vec \pi \,'\cdot \vec \varepsilon }
+
e^{-\frac{i}{\hbar} \vec \pi \cdot \vec \varepsilon }
e^{\frac{i}{\hbar} \vec \pi \,'\cdot \vec \varepsilon }
\Big)
\\
&\!\!\!\!\!\!\!=
|M|^2 \int_{B_R} \frac{d\vec \varepsilon}{\sqrt{R^2-\vec\varepsilon\,^2}}
\;2\; e^{-\frac{i}{\hbar} \vec \varepsilon \cdot (\vec \pi - \vec\pi\,')}
\!= 
4 \pi^2\hbar^2 \kappa |M|^2\frac{ J_1(\kappa\|\vec \pi - \vec\pi\,'\|)}{\|\vec \pi - \vec\pi\,'\|}
=4 \pi^2\hbar^2 \kappa^2 |M|^2 k_1(\kappa\|\vec \pi - \vec\pi\,'\|)
\end{align*}
(see Appendix \ref{AppendixBessel}). Similarly: 
\begin{align*}
\langle \phi_{\vec \pi \bullet} , \phi_{\vec \pi \,' \bullet} \rangle_c 
&=  
4 \pi^2\hbar^4 \kappa^4 |M'|^2 k_2(\kappa\|\vec \pi - \vec\pi\,'\|)\,,
\\
\langle \phi_{\vec \pi \circ} , \phi_{\vec \pi \,' \bullet} \rangle_c
&=  0\,,
\end{align*}
showing that the two families of states are orthogonal to each other and, 
more importantly, that both kinds of ``plane waves'' in configuration space 
$\phi_{\vec \pi \circ}$ and $\phi_{\vec \pi \bullet}$ are normalizable and
non-orthogonal. We can check, taking $\vec\pi = \vec\pi'$, that 
$M=\frac{1}{\sqrt{2}\pi \hbar \kappa}$ 
and $M'=\frac{\sqrt{2}}{\pi \hbar \kappa^2}$ for  normalized $\vec\pi$-states.
Let us also note that we have just found the wave function components of states 
$\phi_{\vec \pi \circ}$ and $\phi_{\vec \pi \bullet}$ in momentum space. Therefore, we
can write: 
\begin{align}
\phi_{\vec \pi '\circ}(\vec\pi)&=(\phicirc_{\vec \pi '\circ}(\vec\pi),\phibullet_{\vec \pi '\circ}(\vec\pi))= (\frac{2}{\kappa} \frac{ J_1(\kappa\|\vec \pi - \vec\pi\,'\|)}{\|\vec \pi - \vec\pi\,'\|},0) = (2 k_1(\kappa\|\vec \pi - \vec\pi\,'\|),0)   \nonumber \\
  \phi_{\vec \pi '\bullet}(\vec\pi)&=(\phicirc_{\vec \pi '\bullet}(\vec\pi),\phibullet_{\vec \pi '\bullet}(\vec\pi))= (0,8 \hbar^2\frac{ J_2(\kappa\|\vec \pi - \vec\pi\,'\|)}{\kappa^2\|\vec \pi - \vec\pi\,'\|^2}) = (0,8 \hbar^2 k_2(\kappa\|\vec \pi - \vec\pi\,'\|))\,. \label{estadosnumomento}
 \end{align}

We put forward the following expansion of configuration wave functions in terms of 
states $\phi_{\vec \pi \circ}$ and $\phi_{\vec \pi \bullet}$, inspired in scalar 
product \eqref{ProductoEscalarMomentos}:
\begin{equation}
	\psi^\pm(\vec \varepsilon)= D\int_{\mathbb{R}^3} d\vec \pi 
	\int_{\mathbb{R}^3} d\vec \pi' \Big(
\psicirc(\vec\pi)
\kappa^2 k_2(\kappa\|\vec \pi -\vec \pi \,'\|)
\phi_{\vec \pi' \circ}^\pm(\vec\varepsilon\,) \; + \; 
\frac{1}{2}\psibullet(\vec\pi)
k_1(\kappa\|\vec \pi -\vec \pi \,'\|)
\phi_{\vec \pi' \bullet}^\pm(\vec\varepsilon\,)
\Big)\,, \label{ResolutionIdentityMom}
\end{equation}
%
valid for wave function components of states 
$\psi(\vec \varepsilon)\in \mathcal H_c$, where the coefficients of the expansion 
$\psicirc(\vec\pi)$ and $\psibullet(\vec\pi)$ are the wave function components of 
the corresponding states in momentum space and $D=\frac{\kappa^3\sqrt{\hbar}}{8\sqrt{2\pi^3}}$.

Expansion  \eqref{ResolutionIdentityMom} can be seen as a Resolution of the Identity in terms of the \textit{overcomplete} set of states $\phi_{\vec \pi \circ}$ and $\phi_{\vec \pi \bullet}$. Expressed in the terminology of \textit{Frame Theory} \cite{AliAntoineGazeau},  the two sets of states $\phi_{\vec \pi \circ}$ and $\phi_{\vec \pi \bullet}$ jointly define a continuous frame which is tight, i.e. they provide a Resolution 
of the Identity. That means that although they are non-orthogonal among each other, the frame is auto-dual (i.e. it behaves in many respect as if it where an orthogonal basis). This result differs from other constructions like the one used in \cite{Sherman,Volobuyev,WolfWignerFunction-Sphere}, where they introduce a non-orthogonal, discrete-continuous, overcomplete family of states defining a non-tight frame and requiring a \textit{dual frame} to reconstruct any state in terms of them.  

It should be emphasized that both families of states 
$\phi_{\vec \pi \circ}$ and $\phi_{\vec \pi \bullet}$ are required to have a frame, since $\phi_{\vec \pi \circ}$ are even functions  and $\phi_{\vec \pi \bullet}$ are odd functions on $S^3$ (with respect to reflections on the equatorial sphere $S^2$).

Note also that this Resolution of the Identity is unusual in the sense that involves a double integral in the momentum variables with convolutions kernels $k_1$ and $k_2$, the same as the ones appearing in the scalar product \eqref{ProductoEscalarMomentos}.

Expansion  \eqref{ResolutionIdentityMom} can be further simplified: 
\begin{align*}
\psi^\pm(\vec \varepsilon) 
&= \\
&= 
D\int_{\mathbb{R}^3} d\vec \pi 
	\int_{\mathbb{R}^3} d\vec \pi' \Big(
\psicirc(\vec\pi)
\kappa^2 k_2(\kappa\|\vec \pi -\vec \pi \,'\|) 
\frac{e^{\frac{i}{\hbar} \vec \pi' \cdot \vec \varepsilon }}{\sqrt{2}\pi \hbar \kappa} \; \pm i |\varepsilon_4|\; 
\psibullet(\vec\pi)
k_1(\kappa\|\vec \pi -\vec \pi \,'\|)
\frac{e^{\frac{i}{\hbar} \vec{\pi}'\cdot \vec{\varepsilon}}}{\sqrt{2}\pi \hbar^2\kappa^2}  
\Big)
\\
&=
\frac{D}{\sqrt{2}\pi \hbar\kappa}
\int_{\mathbb{R}^3} d\vec \pi 
\Big(
\Big(
\int_{\mathbb{R}^3} d\vec \pi' 
\kappa^2 k_2(\kappa\|\vec \pi -\vec \pi \,'\|)
e^{\frac{i}{\hbar} \vec \varepsilon \cdot (\vec \pi\,' -\vec \pi) }
\Big)
 \psicirc(\vec \pi)
\\
&\pm
i \frac{|\varepsilon_4|}{\hbar \kappa}
\Big(
\int_{\mathbb{R}^3} d\vec \pi' 
k_1(\kappa\|\vec \pi -\vec \pi \,'\|)
e^{\frac{i}{\hbar} \vec \varepsilon \cdot (\vec \pi\,' -\vec \pi) }
\Big)
\psibullet(\vec \pi)
\Big)e^{\frac{i}{\hbar} \vec \varepsilon \cdot \vec \pi }
\,.
\end{align*}
Using again formulas from Appendix \ref{AppendixBessel} (recall that 
$(|\varepsilon_4|)_+^\alpha = |\varepsilon_4|^\alpha$ if $\vec\varepsilon \in B_R$  and 
$(|\varepsilon_4|)_+^\alpha =0$ if $\vec\varepsilon \notin B_R$), we find: 

\begin{equation}
\label{inverseFTbuena}
	\psi^\pm(\vec \varepsilon) =
	\frac{1}{\sqrt{2}(2\pi\hbar)^{3/2}}
\int_{\mathbb{R}^3} d\vec \pi
\Big(
(|\varepsilon_4|)_+ \psicirc(\vec \pi) 
\pm i\hbar (|\varepsilon_4|)_+^0
\psibullet(\vec \pi)\Big)
 e^{\frac{i}{\hbar} \vec \varepsilon \cdot \vec \pi } =\frac{1}{\sqrt{2}} \left( (|\varepsilon_4|)_+\hat{\psicirc}(\vec \varepsilon)\pm i\hbar (|\varepsilon_4|)_+^0
\hat{\psibullet}(\vec \varepsilon)\right)
 \,, 
\end{equation} 
where  $(|\varepsilon_4|)_+^0$ is nothing but the indicator function for $B_R$, so 
that this expression provides non-zero wave function components in configuration 
space only for $\vec\varepsilon \in B_R$. That is the desired generalized  inverse Fourier transformation, recovering expressions \eqref{fourier3}-\eqref{fourier4}.

Inserting \eqref{fourier1} and \eqref{fourier2} in \eqref{inverseFTbuena}, it can be
checked that the composition of the generalized Fourier transform with its inverse is 
the identity in configuration space. Using \eqref{inverseFTbuena} in 
\eqref{fourier1} and \eqref{fourier2} one expects to obtain the identity in 
momentum space, but this is only true for the subspace of \textit{oscillatory} 
solutions of the Helmholtz equation ${\cal V}_m$. In fact, in performing the 
composition, operator 
$\hat{K}_{\frac{3}{2}}$  arises (see Appendix \ref{AppendixBessel}), which projects onto 
${\cal V}_m$. Hence, the generalized Fourier Transform is unitary and invertible 
within its domain of definition. 

One final comment is in order: we see that the scalar 
product in momentum space (involving a double integral, non-local) is in fact 
``diagonalized''  by our generalization of Fourier Transform from momentum space to 
configuration space, where a ``local'', or Lebesgue, scalar product is defined. 
Convolution with kernels 
$k_1(\kappa\|\vec \pi -\vec \pi \,'\|)$ and $k_2(\kappa\|\vec \pi -\vec \pi \,'\|)$ 
in the scalar product becomes multiplicative leaving the scalar product with the 
usual single integration in configuration space and turning ${\cal H}_{\rm m}$ into a $L_\mu^2$ Hilbert space. The 
interested reader can also check  Appendices \ref{DiracNotation} and
\ref{AppendixBessel} for details.

\section{Time evolution in momentum space}
\label{TimeEvolutionMomentum}

Now we turn to the dynamics of the quantum free particle moving on $S^3$ in momentum 
space. As in configuration space, the Hamiltonian is given by 
$\hat{H}\equiv \frac{1}{2m}\delta^{ij}\hat{k}_{mi}\hat{k}_{mj}$ and, in 
principle, we have to look for simultaneous eigenstates of the set of mutually 
commuting operators $\langle \hat{H},\,\hat{J}^2,\,\hat{J}_{3}\rangle$ in order 
to get the stationary states in momentum space. However, if we want an explicit 
expression for the corresponding wave functions, Helmholtz equation must be satisfied 
as well, which amounts to the condition of wave functions being eigenstates of the 
Laplacian in 4D, $\Delta_\pi$ in \eqref{EcHelmholtz}, with eigenvalue 
$\kappa^2=\frac{R^2}{\hbar^2}$ (note that $\Delta_\pi$ commutes with any operator
belonging to  the algebra of $\tilde\Sigma(SU(2))$). For the actual computations, it 
is convenient to resort to hyperspherical $\pi$-variables:  
\begin{align}
\pi_1 &=\,\pi_r \sin{\pi_\chi}\,\sin{\pi_\theta}\,\cos{\pi_\phi}\nn\\
\pi_2 &=\,\pi_r \sin{\pi_\chi}\,\sin{\pi_\theta}\,\sin{\pi_\phi} \label{nuhypersphericalcoor}\\
\pi_3 &=\,\pi_r \sin{\pi_\chi}\,\cos{\pi_\theta}\nn\\
\pi_4 &=\, \pi_r \cos{\pi_\chi}\,,\nn
\end{align}
in which $\hat{H}$ acquires the form: 
\[
\hat{H} = - \frac{1}{2 m \kappa^2}\left( 
\frac{\partial^2}{\partial \pi_\chi^2} + 2\cot{\pi_\chi}\frac{\partial\,}{\partial \pi_\chi}
+ \csc^2{\pi_\chi} \Big( \frac{\partial^2}{\partial \pi_\theta^2} + \cot{\pi_\theta}\frac{\partial\,}{\partial \pi_\theta}
+ \csc^2{\pi_\theta} \frac{\partial^2}{\partial \pi_\phi^2}\Big)
\right)\,.
\]
We see that $\hat{H}$ does not depend on $\pi_r$. In fact, the Hamiltonian in
momentum space turns out to have the same \textit{functional} form as in configuration 
space, 

\begin{equation}
\hat{H} =  - \frac{1}{2 m \kappa^2} \Delta_{mL-B}\,,	
\label{HamMom}
\end{equation}
where $\Delta_{mL-B}$ is the Laplace-Beltrami operator associated with a sphere $S^3$ 
in $\pi$-variables in momentum space. From there, the time-dependent Schr\"odinger 
equation in momentum representation for time-dependent wave functions is given by:

\begin{equation}
 i\hbar \frac{\partial\,}{\partial t}\phi(\vec{\pi},\pi_4,t)=- \frac{1}{2 m \kappa^2} \Delta_{mL-B}\,\phi(\vec{\pi},\pi_4,t)\,.
 \label{ChorriMom}
\end{equation}

The well-known formula for the Euclidean Laplacian in (hyper-)spherical coordinates 
allows to write the Helmholtz equation in a convenient form: 
\[
\left[
\frac{\partial^2}{\partial \pi_r^2} + \frac{3}{\pi_r}\frac{\partial\,}{\partial \pi_r}+\frac{1}{\pi_r^2}\Delta_{mL-B}+\kappa^2\right]\phi(\vec{\pi},\pi_4)=0\,.
\]

Eigenfuctions of the Hamiltonian will have a functional form in terms of angular 
$\pi$-variables similar to that in configuration space in terms of angular 
$\varepsilon$-variables, but now the dependence on $\pi_r$ is chosen so that 
eigenfunctions solve the Helmholtz equation. 

The expressions of $\hat{J}^2$ and $\hat{J}_{3}$ are given by: 
\[
\hat{J}^2 = \hbar^2\left(\frac{\partial^2}{\partial \pi_\theta^2} + \cot{\pi_\theta}\frac{\partial\,}{\partial \pi_\theta}
+ \csc^2{\pi_\theta} \frac{\partial^2}{\partial \pi_\phi^2}\right)\,, \qquad 
\hat{J}_{m3} = \hbar\frac{\partial\,}{\partial \pi_\phi}\,.
\]

Stationary states $\phi_{nlm}(\pi_r,\pi_\chi,\pi_\theta,\pi_\phi)$ satisfying the 
Helmholtz equation can be found: 

\begin{equation}
	\phi_{nlm}(\pi_r,\pi_\chi,\pi_\theta,\pi_\phi)=M_{nlm}\sin^l{\pi_\chi}\,C^{(l+1)}_{n-l}(\cos{\pi_\chi})Y_{lm}(\pi_\theta,\pi_\phi) \big(A \,J_{n+1}(\kappa \pi_r)+B \,Y_{n+1}(\kappa \pi_r)\big), \label{wavefunctionsmom}
\end{equation}
where $C^{(l+1)}_{n-l}(x)$ are the Gegenbauer polynomials, 
$Y_{lm}(\pi_\theta,\pi_\phi)$ are the spherical harmonics in $\pi$-variables, 
$J_\alpha(z)$ and $Y_\alpha(z)$ are the Bessel functions of the first and second kind, 
respectively, and $M_{nlm}$, $A$ and $B$ are normalizing constants to be determined. They solve the 
eigenvalue equations for $\langle \hat{H},\,\hat{J}^2,\,\hat{J}_{3}\rangle$: 
\begin{align*}
\hat{H}\phi_{nlm}&=n(n+2)\frac{1}{2 m \kappa^2}\phi_{nlm}\\
\hat{J}^2\phi_{nlm}&=l(l+1)\hbar^2\phi_{nlm}\\
\hat{J}_{3}\phi_{nlm}&=m\, \hbar \,\phi_{nlm}\,.
\end{align*}

The normalizing constants $M_{nlm}$ can be found directly in momentum space using 
$\langle\cdot,\cdot\rangle_m$. See Appendix \ref{Mmomentos} for the detailed computation. In particular it 
is proven that $n, l, m$ take the standard integer values with the restriction $n\geq l\geq 0$ and $|m|\leq l$.

\section{Conclusions and final remarks}

 In this paper we have fully developed the momentum space quantization of a particle moving on the sphere $S^3$. The starting point is the  group of (contact) symmetries of the Poincar\'e-Cartan form associated
with the free motion on $S^3$ (a subgroup of the Euclidean group $E(4)$ for a spin-less particle), which 
characterizes the system and that was derived in \cite{SU(2)} along with the quantization in configuration space, which only required the use of a (standard) first-order polarization. There the momentum space was also briefly discussed, providing the (new) higher-order polarization required for its derivation.  Here we have soundly developed the Hilbert space in momentum representation, which involves a non-local positive-definite invariant scalar product (with an integral convolution kernel made of Bessel functions), and have constructed the Fourier Transform relating (unitarily) the momentum and configuration spaces, which is also non-trivial and differs from the usual one in the flat case. This construction is similar to the one describing Helmholtz Optics in 4D (see \cite{BernardoOptica} for the case of 3D), but with the roles of configuration and momentum spaces interchanged, and the fact that the former is a quantum system and the latter is a classical field.

Another important result of this paper is the definition of an overcomplete set (frame) of states labelled by the (continuous) momenta which are normalizable and non-orthogonal, providing a Resolution of the Identity (tight frame). This can be compared with other families of states, like the one introduced in \cite{Sherman,Volobuyev,WolfWignerFunction-Sphere} generalizing Fourier Series of the  $S^1$ case, with normalizable and non-orthogonal states labelled by a ``momentum'' with discretized norm, which does not provide a Resolution of the Identity (non-tight frame), thus requiring a dual family (frame) for reconstruction of states.

%
%

Once the Hilbert space in momentum space has been constructed, the time evolution under the Hamiltonian associated with the free (geodesic) motion is described, using hyperspherical coordinates. Although the Hamiltonian does not close a finite-dimensional Lie algebra with the  algebra of basic symmetries, a complete description can be given constructing the eigenvalues of the Hamiltonian and its common eigenstates with respect to a suitable set of mutually commuting operators. These eigenstates are checked to be orthonormal with respect to the non-local scalar product in momentum space.


The study made in this paper for the case of $S^3$ can be easily generalized to any sphere $S^n$ with $n\geq 1$, with the only difference that the symmetry group will be the full Euclidean group
$E(n+1)$ (in the case discussed here of $n=3$ the symmetry group is smaller in the spin-less case). The scalar product in momentum space will be non-local, with integral convolution kernels
made of appropriate Bessel functions. All other results given in this paper generalize straightforwardly to the general case, and will be published elsewhere.

The representation in momentum space for a particle moving on the sphere $S^3$ given here is, as far as we know, new in the quantum mechanical context. There are other descriptions, that generalize the Fourier Series description in the case of $S^1$, like the Sherman-Volobuyev basis \cite{WolfWignerFunction-Sphere}, where the momentum has a discretized norm but its direction can be arbitrary. In our case, the momentum $(\vec\pi,\pi_4)$ lies in $\mathbb{R}^4$ and is therefore continuous, i.e. the momentum can be seen as the one of the ambient (Euclidean) space where the sphere is embedded (the constraint is imposed at the level of the wavefunctions through the Helmholtz equation).

The construction of coherent states for a particle on the sphere $S^n$ \cite{deBievre,deBievreGonzalez,Kowalski,MOlmoCS,Kastrup,GazeauPLA} or for its associated symmetry group $E(n+1)$ \cite{IshamKlauder} is a long-standing problem and still 
today there are contributions to it \cite{GazeauCS}. A construction of a family of CS for the Euclidean group relying on the basis given by Eq. \eqref{estadosnuconfig} (or \eqref{estadosnumomento} on momentum space) and the scalar product on momentum space with nice properties will be given elsewhere.

Also, the introduction of an appropriate Wigner function for a particle on the sphere (and other curved spaces) has attracted much attention \cite{WolfWignerFunction-Sphere,WignerFunctionEuclieo,WignerHelmholtz}. The definition of a Wigner function using the momentum representation 
introduced here will also be given elsewhere.

It is also interesting that results from signal analysis and sampling theory help us to reconciliate the standard, discrete image of the momentum and the continuous image of the momentum discussed in this paper. The key point is that the Hilbert space in momentum space is made of functions whose Fourier spectrum is bounded (\textit{band limited} in the signal analysis jargon). In that case, the 
functions can be reconstructed from a set of (infinite) discrete values using sinc-type functions (see Appendix \ref{SpacesLk}). This picture was first introduced in \cite{WignerFunctionEuclieo} for the Wigner function and later developed in \cite{SamplingHelmholtz,MeasureingHelmholtz,KastrupWigner1,KastrupWigner2}. The use of this ideas for a similar relation between the discrete picture and the continuous picture of the momentum in the context of the quantum mechanical particle on the sphere will also be given elsewhere.

\appendix

\section{Configuration space quantization}

\label{ReviewConfiguration}

We review in this Appendix the main results given in \cite{SU(2)} providing some more technical insight, required to parallel the results given in momentum space quantization for the free particle moving freely on a $S^3$ sphere. We start (see Section~\ref{Configuracion}) by recalling that for the Lie algebra corresponding to the Lie group $\tilde\Sigma(SU(2))$, symmetry of our physical system, a polarization, i.e.
a maximal left sub-algebra, containing the characteristic sub-algebra and excluding the central generator, exists:
\begin{equation}
\mathcal{P}_{\rm c}=\langle Z_{(\vec\pi)}^{L},\, Z_{(\pi_4)}^{L}\rangle\,,
\end{equation}
leading to  wave functions on ``configuration space'' by imposing $\mathcal{P}_{\rm c}\varPsi(\zeta,\vec{\varepsilon},\vec{\pi},\pi_4)=0\,$. For convenience, let us firstly perform
the following redefinition in the wave functions:
\begin{equation}
 \varPsi(\zeta,\vec{\varepsilon},\vec{\pi},\pi_4)=\zeta e^{i\kappa \pi_4}\psi(\vec{\varepsilon},\vec{\pi},\pi_4)\,.
\end{equation}

Then, the polarization equations leads to:

\begin{equation}
\psi^\pm(\vec{\varepsilon},\vec{\pi},\pi_4)=\zeta
e^{-i\frac{m }{\hbar}(\vec{\varepsilon}\cdot\vec{\pi}\pm |\varepsilon_4|\pi_4)}\phi^\pm(\vec{\varepsilon})\,,
\end{equation}
where the $\pm$ signs correspond to the expression of $\psi$ in each of the local charts $U_\pm$. 
The quantum operators given by the right-invariant generators restricted to the subspace of polarized functions are:
\bea
\hat{\vec k}\phi^\pm(\vec{\varepsilon})&=& i\hbar Z^R_{(\vec\varepsilon)} \phi^\pm(\vec{\varepsilon}) = - \frac{i}{\kappa}\left(\pm |\varepsilon_4| \frac{\partial\,}{\partial\vec\varepsilon} + \vec\varepsilon\times \frac{\partial\,}{\partial\vec\varepsilon} \right)\phi^\pm(\vec{\varepsilon})\nn\\
\hat{\vec\varepsilon}\phi^\pm(\vec{\varepsilon})&=&  i\hbar Z^R_{(\vec\pi)}\phi^\pm(\vec{\varepsilon})   = \vec\varepsilon\,\phi^\pm(\vec{\varepsilon}) \label{PincheRed}\\
 \hat{\varepsilon}_4\phi^\pm(\vec{\varepsilon})&=&     i\hbar Z^R_{(\pi_4)} \phi^\pm(\vec{\varepsilon}) = \pm |\varepsilon_4|\phi^\pm(\vec{\varepsilon}) \equiv \pm R\,\hat{\rho}\, \phi^\pm(\vec{\varepsilon}).\nn  
\eea

The scalar product in configuration space was given in \cite{SU(2)} through the integration measure
\be
d\mu_{\rm c}= 
\frac{1}{|\varepsilon_4|}\, d\varepsilon^{1}\wedge d\varepsilon^{2}\wedge d\varepsilon^{3}\,.
\label{measureconf}
\ee

With this integration measure the scalar product is:
\begin{equation}
 \langle \phi, \phi'\rangle_{\rm c}= \int_{B_R}\frac{d\vec\varepsilon}{|\varepsilon_4|}
\left(\phi^+(\vec\varepsilon)^*\phi'{}^+(\vec\varepsilon)+\phi^-(\vec\varepsilon)^*\phi'{}^-(\vec\varepsilon)\right)\,, \label{scalarproductconf}
\end{equation}
where $\phi^\pm,\phi'{}^\pm$ denote the local expressions of $\phi,\phi'$ in $U_\pm$.

The Hilbert space $\Hc$ is defined as the completion with respect to this scalar product of the 
subspace of normalizable couples $(\phi^+(\vec\varepsilon),\phi^-(\vec\varepsilon))$.

The operators (\ref{PincheRed}) realize a unitary and irreducible representation of the group $\tilde G$ on $\Hc$.
This representation can be extended \cite{SU(2)} to a unitary and irreducible representation of the Euclidean group $E(4)$, with the addition of the operator $\hat{\vec J}$:
\begin{equation}
 \hat{\vec J}\phi^\pm(\vec{\varepsilon})=   - i\hbar  \left( \vec\varepsilon\times \frac{\partial\,}{\partial\vec\varepsilon}\right) \phi^\pm(\vec{\varepsilon})\,.
\end{equation}

Although the time evolution was disregarded in the quantization process, it comes now in a natural way as the Hamiltonian proves to be 
unambiguously defined in terms of basic operators: $\hat{H}\equiv\frac{1}{2m}\delta^{ij}\hat{k}_i\hat{k}_j$ reproducing the expected expression:
\be
\hat{H}\phi^\pm(\vec{\varepsilon})=-\frac{\hbar^2}{2m}\Delta_{\text{L-B}}\phi^\pm(\vec{\varepsilon})=
-\frac{m}{2\kappa^2}\left(-3\varepsilon^{j}\frac{\partial\,}{\partial\varepsilon^{j}}+(R^2\delta^{kj}-
\varepsilon^{k}\varepsilon^{j})\frac{\partial^{2}\,}{\partial\varepsilon^{k}\partial\varepsilon^{j}}\right)\phi^\pm(\vec{\varepsilon})\,,
\ee
where $\Delta_{\text{L-B}}$ stands for the Laplace-Beltrami operator associated with the metric in the classical Lagrangian.

The evolution in time of the state $\phi(\vec{\varepsilon},t)$ is given by the Schr\"odinger equation:
\begin{equation}
 i\hbar \frac{\partial\,}{\partial t}\phi^\pm(\vec{\varepsilon},t)=\hat{H}\phi^\pm(\vec{\varepsilon},t)
 \label{ChorriConf}
\end{equation}

\subsection{Bases in configuration space representation}

Let us introduce some convenient bases in the Hilbert space $\Hc$ which will be useful to expand an arbitrary state in terms of them. These bases will be obtained as eigenstates of  sets of commuting operators.

\subsubsection{Basis of position eigenstates}

The first basis will be that of common eigenstates of the commuting  operators $\hat{\vec\varepsilon}$. The spectrum of these operators are continuous and doubly degenerated, therefore the associated eigenfunctions will be distributions:
\begin{eqnarray}
  \phi_{\vec{\varepsilon}\,'+}(\vec\varepsilon)&=&(\phi_{\vec{\varepsilon}\,'+}^{+}(\vec\varepsilon),\phi_{\vec{\varepsilon}\,'+}^{-}(\vec\varepsilon))= (|\varepsilon_4|\delta^{(3)}(\vec\varepsilon-\vec{\varepsilon}\,'), 0)  \nonumber \\
  \phi_{\vec{\varepsilon}\,'-}(\vec\varepsilon)&=&(\phi_{\vec{\varepsilon}\,'-}^{+}(\vec\varepsilon),\phi_{\vec{\varepsilon}\,'-}^{-}(\vec\varepsilon))= (0, -|\varepsilon_4|\delta^{(3)}(\vec\varepsilon-\vec{\varepsilon}\,')) \,, \label{PositionEigenstatesConf}
\end{eqnarray}
with $\vec\varepsilon\,'\in B_R$.
The state $\phi_{\vec{\varepsilon}\,'+}$ represents a localized particle at the point $(\vec\varepsilon\,',\varepsilon_4')$ with $\varepsilon_4'=\sqrt{R^2-\vec{\varepsilon}\,'{}^2}>0$, and $\phi_{\vec{\varepsilon}\,'-}$ represents a localized particle at the point $(\vec\varepsilon\,',\varepsilon_4')$ with $\varepsilon_4'=-\sqrt{R^2-\vec{\varepsilon}\,'{}^2}<0$.

Note that the signs $\pm$ in the subscripts correspond to different globally defined functions on the sphere $S^3$,
whereas in the superscript indicate two different local expressions of the same object. In the following,
to avoid confusion, a $\pm$ sign in a superscript will always mean a local chart expression of a function, whereas a $\pm$ sign in a subscript means a global
expression of a function, or the associated Hilbert spaces and projection operators.

There is also a nondegenerate part of the spectrum, when $\|\vec\varepsilon\,'\|=R$, and therefore $\varepsilon_4'=0$, corresponding to states $\phi_{\vec{\varepsilon}\,'0}$ localized at the equatorial sphere $S^2$.

Let us define the Hilbert subspaces $\Hc{}_{(\pm)}$ as those expanded by the states $\phi_{\vec{\varepsilon}\,'\pm}$, with $\vec\varepsilon\,'\in B_R$, respectively. Clearly these subspaces are orthogonal with respect to the scalar 
product (\ref{scalarproductconf}).  $\Hc{}_{(+)}$ is made of wave functions with support on the \textit{northern} hemisphere of $S^3$ (with $\varepsilon_4>0$), whereas $\Hc{}_{(-)}$ is made of wave functions with support on 
the \textit{southern} hemisphere of  $S^3$ (with $\varepsilon_4<0$).
Let us define also $\Hc{}_{(0)}$ as the Hilbert space expanded by the states $\phi_{\vec{\varepsilon}\,'0}$, with $\vec\varepsilon\,'\in S^2$. $\Hc{}_{(0)}$ is made of wave functions with support on the equatorial sphere $S^2$.

We have that
\begin{equation}
 \Hc = \Hc{}_{(+)} \oplus \Hc{}_{(0)} \oplus \Hc{}_{(-)}\,,\qquad
 \Hc = \overline{\Hc{}_{(+)} \oplus \Hc{}_{(-)}}
\end{equation}

The last equation holds since $\Hc{}_{(0)}$  has zero measure with respect to the measure (\ref{measureconf}), and therefore can be discarded when expanding a general state in terms of eigenstates of $\hat{\vec\varepsilon}$.

Note also that the states
localized near the equatorial sphere $S^2$, with $\|\vec\varepsilon\,'\|\lesssim R$ and $|\varepsilon_4'|\approx 0$, are ``supressed'' by the factor $|\varepsilon_4|$ in the wave functions (\ref{PositionEigenstatesConf}), but this is compensated by the integration measure (\ref{measureconf}).

We can define the orthogonal projectors $\mathbb{P}_c{}_\pm$ projecting onto the subspaces $\Hc{}_{(\pm)}$. Define $\phi{}_{(\pm)}=\mathbb{P}_c{}_\pm\phi$ $\forall \phi = (\phi^+,\phi^-)\in \Hc$. Then $\phi{}_{(+)}=(\phi^+,0)$ and $\phi{}_{(-)}=(0,\phi^-)$.

An arbitrary state $\phi\in \Hc$ can be expanded in terms of the states $\phi_{\vec{\varepsilon}\,',\pm}$, with $\vec\varepsilon\,'\in B_R$:
\begin{equation}
\phi(\vec\varepsilon)=\int_{B_R}\frac{d\vec\varepsilon\,'}{|\varepsilon_4'|}
\left(\Phi{}_+(\vec\varepsilon\,')  \phi_{\vec{\varepsilon}\,'+}(\vec\varepsilon)  + \Phi{}_-(\vec\varepsilon\,')  \phi_{\vec{\varepsilon}\,'-}(\vec\varepsilon)\right)\,,
\label{ResolutionIdentityConf2}
\end{equation}
with $\Phi{}_\pm(\vec\varepsilon\,')=\langle \phi_{\vec{\varepsilon}\,'\pm},\phi\rangle_{\rm c}$, i.e. we can write $\phi^\pm(\vec\varepsilon)=\Phi{}_\pm(\vec\varepsilon)$. See below for a proof of this statement.

Using Dirac's bra-ket notation (see Appendix \ref{DiracNotation} for a review), we introduce the generalized \textit{ket} $|\vec\varepsilon\pm\rangle$ associated with the state $\phi_{\vec{\varepsilon}\pm}$,
in such a way that these states generate the representation in configuration space in the sense
that $\phi^\pm(\vec\varepsilon)=  \langle \phi_{\vec{\varepsilon}\pm},\phi\rangle_{\rm c}\equiv  \langle \vec{\varepsilon}\pm|\phi\rangle$, and the states $|\vec\varepsilon\pm\rangle$ satisfy a resolution of the identity:
\begin{equation}
 I_{\Hc} = \int_{B_R}\frac{d\vec\varepsilon}{|\varepsilon_4|}\left( |\vec\varepsilon+\rangle \langle \vec\varepsilon\!+\!|\, +\, |\vec\varepsilon-\rangle \langle \vec\varepsilon\!-\!|
\right)\,.\label{ResolutionIdentityConf}
\end{equation}

The proof of (\ref{ResolutionIdentityConf2}) (and (\ref{ResolutionIdentityConf})) can be derived from the spectral theorem for bounded self-adjoint operators and we shall sketch it here. Define the operator
\begin{equation}
 \hat{A}=\int_{B_R}d\vec\varepsilon \mu(\vec\varepsilon)\left( |\vec\varepsilon+\rangle \langle \vec\varepsilon\!+\!|\, +\, |\vec\varepsilon-\rangle \langle \vec\varepsilon\!-\!|\right)
\end{equation}
where $\mu(\vec\varepsilon)$ is a measure density (the spectral measure) to be determined. By construction, $\hat{A}$ is Hermitian. Using that 
\begin{equation}
 \langle \phi_{\vec{\varepsilon}\,'\sigma'},\phi_{\vec{\varepsilon}\sigma}\rangle_{\rm c} =|\varepsilon_4|\delta_{\sigma\sigma'}\delta^3(\vec{\varepsilon}\,'-\vec{\varepsilon})\,,
 \label{overlapposicion}
\end{equation}
with $\sigma,\sigma'=\pm$, we check that $\hat{A}$ is an orthogonal projector if and only if $\mu(\vec\varepsilon)=\frac{1}{|\varepsilon_4|}$. Suppose now that the state $\phi'\in\Hc$ is orthogonal
to $\hat{A}\phi$ for all $\phi\in\Hc$, then
\begin{eqnarray}
 0=\langle \phi',\hat{A}\phi\rangle_{\rm c} &=&\int_{B_R}\frac{d\vec\varepsilon}{|\varepsilon_4|}\left( \langle \phi'|\vec\varepsilon+\rangle \langle \vec\varepsilon\!+\!|\phi\rangle 
 \,+\, \langle \phi'|\vec\varepsilon-\rangle \langle \vec\varepsilon\!-\!|\phi\rangle
\right)\nonumber \\
&=& \int_{B_R}\frac{d\vec\varepsilon}{|\varepsilon_4|}\left(  \phi'{}^+(\vec\varepsilon)^*  \phi^+( \vec\varepsilon) 
 +  \phi'{}^-(\vec\varepsilon)\phi^-( \vec\varepsilon)\right)
\end{eqnarray}
and this implies that $\phi'{}^{\pm}(\vec\varepsilon)=0$, and therefore $\phi'=0$ since the scalar product (\ref{scalarproductconf}) is nondegenerate on $\Hc$. This proves that 
$\hat{A}=I_{\Hc}$.

Let us introduce, for later convenience, the Hilbert subspaces $\Hc{}_{(e)}$ and $\Hc{}_{(o)}$. $\Hc{}_{(e)}$ is made of functions on the sphere $S^3$ satisfying 
$\phi^-=\phi^+$, and $\Hc{}_{(o)}$ is made of functions satisfying $\phi^-=-\phi^+$. That is, $\Hc{}_{(e)}$ is made of even  functions and $\Hc{}_{(o)}$ is made of odd functions on $S^3$, with respect to reflection on the equatorial sphere $S^2$. Clearly these subspaces are orthogonal, and therefore we can write:
\begin{equation}
\Hc = \overline{\Hc{}_{(e)} \oplus \Hc{}_{(o)}}=\overline{\Hc{}_{(e)} }\oplus \Hc{}_{(o)}\,,
\end{equation}
where $\overline{\Hc{}_{(e)} }$ is the completion of $\Hc{}_{(e)}$, that is, including $\Hc{}_{(0)}$.

Let us define several operators related with those subspaces. The operator realizing the reflection with respect to the equatorial sphere $S^2$ is
given by:
\begin{equation}
 \hat{R} = \int_{B_R}\frac{d\vec\varepsilon}{|\varepsilon_4|}\left( |\vec\varepsilon-\rangle \langle \vec\varepsilon\!+\!| \,+\, |\vec\varepsilon+\rangle \langle \vec\varepsilon\!-\!|
\right)\,.
\end{equation}

Define the orthogonal projectors $\mathbb{P}^{e}_c$ and $\mathbb{P}^{o}_c$ projecting into the subspaces $\Hc{}_{(e)}$ and $\Hc{}_{(o)}$, respectively. Then we have that 
\begin{equation}
 \mathbb{P}^{e}_c = \frac{1}{2}\left(I_{\Hc}+ \hat{R} \right)\,,\qquad \mathbb{P}^{o}_c = \frac{1}{2}\left(I_{\Hc}- \hat{R} \right)
\end{equation}

Since $\varepsilon_4$ is an odd function on $S^3$, then $\hat{\varepsilon}_4\Hc{}_{(e)}\subset \Hc{}_{(o)}$ and $\hat{\varepsilon}_4\Hc{}_{(o)}\subset \Hc{}_{(e)}$, with

\begin{equation}
 \hat{\varepsilon}_4= \int_{B_R}d\vec\varepsilon\left( |\vec\varepsilon+\rangle \langle \vec\varepsilon\!+\!|\, -\, |\vec\varepsilon-\rangle \langle \vec\varepsilon\!-\!|
\right)\,.
\end{equation}

Let us introduce the \textit{sign} operator $\hat{s}$ as:
\begin{equation}
 \hat{s}= \int_{B_R}\frac{d\vec\varepsilon}{|\varepsilon_4|} \left( |\vec\varepsilon+\rangle \langle \vec\varepsilon\!+\!| \,-\, |\vec\varepsilon-\rangle \langle \vec\varepsilon\!-\!|
\right)= \mathbb{P}^{+}_c-\mathbb{P}^{-}_c\,.
\end{equation}

This operator has as eigenspaces $\Hc^{(+)}$, with eigenvalue $+1$, $\Hc^{(-)}$, with eigenvalue $-1$, and $\Hc^{(0)}$, with eigenvalue $0$.
Note that $\hat{s}\Hc^{(e)}= \Hc^{(o)}$ and $\hat{s}\Hc^{(o)}= \Hc^{(e)}$.

Finally, defining the operator 
\begin{equation}
 \widehat{|\varepsilon_4|}= \int_{B_R}d\vec\varepsilon\left( |\vec\varepsilon+\rangle \langle \vec\varepsilon\!+\!|\, +\, |\vec\varepsilon-\rangle \langle \vec\varepsilon\!-\!|
\right)\,,
\end{equation}
we have that $\hat{\varepsilon}_4=\hat{s} \widehat{|\varepsilon_4|}=  \widehat{|\varepsilon_4|}\hat{s}$.

It is also possible to define the even and odd states 
\begin{equation}
 \phi_{\vec{\varepsilon}\pm}^{e}=\frac{1}{2}\left(\phi_{\vec{\varepsilon}+}+\phi_{\vec{\varepsilon}-}\right)\,,\qquad
 \phi_{\vec{\varepsilon}\pm}^{o}=\pm\frac{1}{2}\left(\phi_{\vec{\varepsilon}+}-\phi_{\vec{\varepsilon}-}\right)
 \label{evenoddconf}
\end{equation}
respectively. Note that $\phi_{\vec{\varepsilon}-}^{e}=\phi_{\vec{\varepsilon}+}^{e}
\equiv \phi_{\vec{\varepsilon}}^{e}$ and $\phi_{\vec{\varepsilon}-}^{o}=-\phi_{\vec{\varepsilon}+}^{o}\equiv - \phi_{\vec{\varepsilon}}^{o}$.
These states correspond to  particles localized simultaneously at points symmetrical with 
respect to the equatorial sphere $S^2$ (even and odd Schr\"odinger's cat states).

It should be noted that the continuous basis $\phi_{\vec{\varepsilon}\pm}$ constructed here as eigenstates of 
the commuting  operators $\hat{\vec\varepsilon}$ is the $S^3$ version of the continuous bases on $S^1$ and $S^2$ introduced in \cite{MOlmo1} and \cite{MOlmo2}, respectively. Note, however, that our construction based on the $\tilde\Sigma(SU(2))$ 
group (or the Euclidean group) is more natural and rigorous, since we do not resort to the ill-defined operators ``multiplication by an angle'' \cite{MOlmo2}, but to the Lie algebra generators $\hat{\vec\varepsilon}$.

\subsubsection{Basis of  eigenstates of the Hamiltonian}

The next basis will be that of common eigenstates of the commuting  operators $\langle
\hat{H},\,\hat{J}^2,\,\hat{J}_3\rangle$. It is convenient to resort to hyperspherical coordinates (\ref{hypersphericalcoor}) where the required eigen-problem can be easily solved
with the result \cite{SU(2)}:
\be
\psi_{nlm}(\chi,\theta,\phi)=N_{nl}(\sin\chi)^l\,C^{(l+1)}_{n-l}(\cos{\chi})Y_{lm}(\theta,\phi), \label{wavefunctions}
\ee
where $C^{(l+1)}_{n-l}(x)$ are the Gegenbauer polynomials in the $x$ variable, $Y_{lm}(\theta,\phi)$ are the ordinary spherical harmonics, and
$N_{nl}$ are the following normalizing constants:
\be
N_{nl}=2^ll!\sqrt{\frac{2(n+1)(n-l)!}{\pi (n+l+1)!}}\,.\label{normalizationconf}
\ee

The range of the parameters $(n,l,m)$ are: 
\begin{equation}
 n=0,1,\ldots \,,\qquad l=0,1,2,\ldots n \,,\qquad m=-l,\ldots, l\,. \label{nlm}
\end{equation}

The wave functions $\psi_{nlm}(\chi,\theta,\phi)$ solve the eigen-problem according to the expressions:
\bea
\hat{H}\psi_{nlm}&=&\frac{n(n+2)}{2m\kappa^2}\psi_{nlm}\nn\\
\hat{J}^2\psi_{nlm}&=&l(l+1)\hbar^2 \psi_{nlm}\nn\\
\hat{J}_3\psi_{nlm}&=&m\,\hbar\,\psi_{nlm}\,.\nn
\eea

The set of states $\{\psi_{nlm}\}$ constitute an orthonormal basis of $\Hc$ \cite{hyperspherical} with respect to the scalar product\footnote{As commented previously in hyperspherical coordinates only one chart is needed to define the scalar product, i.e. the set of points not covered by the coordinate chart (\ref{hypersphericalcoor}) is of zero measure. However, extra compatibility conditions for changes
of local charts in the intersection of two charts are needed for a wave function $\psi(\chi,\theta,\phi)$ to be well-defined on the sphere $S^3$, and this is why $n$, $l$ and $m$ take the specified values (\ref{nlm}).}
\begin{equation}
\langle \psi,\psi' \rangle_{\rm c} = \frac{1}{4\pi^2}\int_0^{\pi}d\chi(\sin\chi)^2\int_0^{\pi} d\theta\sin\theta\int_0^{2\pi} d\phi \psi(\chi,\theta,\phi)^*\psi'(\chi,\theta,\phi)
\end{equation}

If we denote by $|n,l,m\rangle$ the \textit{normalized} ket associated with the state $\psi_{nlm}$, then we have the resolution of the identity:
\begin{equation}
 \sum_{n=0}^\infty \sum_{l=0}^n \sum_{m=-l}^l |n,l,m\rangle \langle n,l,m| = I_{\Hc}
\end{equation}

\section{Solutions to Helmholtz equation}
\label{AppendixHelmholtz}

In this Appendix we shall obtain explicitly the solution of the IPV for Helmholtz equation \eqref{IVP-Helmholtz}, i.e.
\be
\left[\frac{\partial^2}{\partial\vec{\pi}^2}+\frac{\partial^2}{\partial
\pi_4^2}+\kappa^2\right]\phi(\vec{\pi},\pi_4)=0\,,\qquad \phi(\vec{\pi},0)=\phicirc(\vec{\pi})\,,\quad \frac{\partial\,}{\partial{\pi_4}}\phi(\vec{\pi},0)=\phibullet(\vec{\pi})\,.
\label{IVP-Helmholtz-Appendix}
\ee
and obtain the conditions for this IVP to be well-defined. We shall also determine  subspaces of solutions invariant under the group $\tilde\Sigma(SU(2))$ (and the Euclidean group $E(4)$).

Expanding an arbitrary solution in Fourier components\footnote{Note that here we are using the standard Fourier transform on $\mathbb{R}^3$, see later for a comparison with the Generalized Fourier Transform introduced in Sec. \ref{Fourier} when restricted to $B_R$.} with respect to $\vec{\pi}$:
\begin{equation}
 \phi(\vec\pi,\pi_4)=\int_{\mathbb{R}^3}\frac{d\vec{\varepsilon}}{(2\pi\hbar)^\frac{3}{2} }\,\hat\phi(\vec\varepsilon,\pi_4) e^{\frac{i}{\hbar}\vec\varepsilon\cdot\vec\pi}
\end{equation}
we have that $\hat\phi(\vec\varepsilon,\pi_4)$ satisfies:
\begin{equation}
 \left[\hbar^2\frac{\partial^2}{\partial\pi_4^2}+ \varepsilon_4^2\right]\hat{\phi}(\vec{\varepsilon},\pi_4)=0
\end{equation}
with $\varepsilon_4^2\equiv R^2-\vec{\varepsilon}{\,}^2$. The solutions to this equation can be classified as follows:
\begin{equation}
 \hat\phi(\vec\varepsilon,\pi_4)=\left\{ \begin{array}{lcl}
                                          e^{\frac{i}{\hbar}|\varepsilon_4|\pi_4} \phi_+(\vec\varepsilon) + e^{-\frac{i}{\hbar}|\varepsilon_4|\pi_4} \phi_-(\vec\varepsilon)  & {\rm if} & \|\vec\varepsilon\|<R \quad (0<\varepsilon_4^2<R^2)\\
                                          \phi_\circ(\vec\varepsilon) + \pi_4 \phi_\bullet(\vec\varepsilon) & {\rm if} & \|\vec\varepsilon\|=R \quad (\varepsilon_4^2=0) \\
                                          e^{\frac{1}{\hbar}|i\varepsilon_4|\pi_4} \varphi_+(\vec\varepsilon) + e^{-\frac{1}{\hbar}|i\varepsilon_4|\pi_4} \varphi_-(\vec\varepsilon)  & {\rm if} & \|\vec\varepsilon\|>R \quad (\varepsilon_4^2<0)
                                         \end{array}
                                         \right.
\end{equation}

Let us denote by $B_R= \{\vec\varepsilon\in\mathbb{R}^3\,/\, \|\vec\varepsilon\|<R \}$, $\bar B_R= \{\vec\varepsilon\in\mathbb{R}^3\,/\, \|\vec\varepsilon\|\leq R \}$, Fr$(B_R)=\{\vec\varepsilon\in\mathbb{R}^3\,/\, \|\vec\varepsilon\|=R \}$ and Ext$(\bar B_R)=\{\vec\varepsilon\in\mathbb{R}^3\,/\, \|\vec\varepsilon\|>R \}$. 

We observe that for $\vec\varepsilon\in B_R$ the solutions are oscillatory in $\pi_4$, for $\vec\varepsilon\in {\rm Fr}(B_R)$ they evolve linearly in $\pi_4$, and for $\vec\varepsilon\in {\rm Ext}(\bar B_R)$ they are exponential in $\pi_4$. Therefore, the only bounded solutions (with respect to $\pi_4$)  are those with $\vec\varepsilon\in B_R$ or $\vec\varepsilon\in {\rm Fr}(B_R)$ but with $\phi_\bullet(\vec\varepsilon)$=0.

Thus a general solution to Helmholtz equation can be expanded as:

\begin{eqnarray}
 \phi(\vec\pi,\pi_4)&=&\int_{B_R}\frac{d\vec{\varepsilon}}{(2\pi\hbar)^\frac{3}{2} }\,e^{\frac{i}{\hbar}\vec\varepsilon\cdot\vec\pi} \left(e^{\frac{i}{\hbar}|\varepsilon_4|\pi_4} \phi_+(\vec\varepsilon) + e^{-\frac{i}{\hbar}|\varepsilon_4|\pi_4}\phi_-(\vec\varepsilon)\right) \nonumber \\ & & + \int_{{\rm Fr}(B_R)}\frac{d\vec{\varepsilon}}{(2\pi\hbar)^\frac{3}{2} }\,e^{\frac{i}{\hbar}\vec\varepsilon\cdot\vec\pi} \left(\phi_\circ(\vec\varepsilon)
  + \pi_4 \phi_\bullet(\vec\varepsilon)\right) \label{FourierExp} \\ & & + \int_{{\rm Ext}(\bar B_R)}\frac{d\vec{\varepsilon}}{(2\pi\hbar)^\frac{3}{2} }\,e^{\frac{i}{\hbar}\vec\varepsilon\cdot\vec\pi} \left(e^{\frac{1}{\hbar}|i\varepsilon_4|\pi_4} \varphi_+(\vec\varepsilon) + e^{-\frac{1}{\hbar}|i\varepsilon_4|\pi_4} \varphi_-(\vec\varepsilon) \right) \nonumber
\end{eqnarray}

Note that since Fr$(B_R)$ is a set of measure zero in $\mathbb{R}^3$, the corresponding solutions will not contribute to the integral (in the Lebesgue sense). However, we shall consider them in order to study the irreducibility of the representation in momentum space. 

Let us now impose the initial conditions in order to solve the IVP \eqref{IVP-Helmholtz-Appendix}. Evaluating \eqref{FourierExp} at $\nu_4=0$ we have:
\begin{eqnarray}
 \phicirc(\vec\pi)&=&\int_{B_R}\frac{d\vec{\varepsilon}}{(2\pi\hbar)^\frac{3}{2} }\,e^{\frac{i}{\hbar}\vec\varepsilon\cdot\vec\pi} \left( \phi_+(\vec\varepsilon) + \phi_-(\vec\varepsilon)\right) \nonumber \\ & & + \int_{{\rm Fr}(B_R)}\frac{d\vec{\varepsilon}}{(2\pi\hbar)^\frac{3}{2} }\,e^{\frac{i}{\hbar}\vec\varepsilon\cdot\vec\pi} \phi_\circ(\vec\varepsilon)
   \label{FourierExpcirc} \\ & & + \int_{{\rm Ext}(\bar B_R)}\frac{d\vec{\varepsilon}}{(2\pi\hbar)^\frac{3}{2} }\,e^{\frac{i}{\hbar}\vec\varepsilon\cdot\vec\pi} \left( \varphi_+(\vec\varepsilon) +  \varphi_-(\vec\varepsilon) \right) \nonumber
\end{eqnarray}

Derivating \eqref{FourierExp} with respect to $\nu_4$ and evaluating  at $\nu_4=0$ we have:

\begin{eqnarray}
 \phibullet(\vec\pi)&=&\frac{i}{\hbar} \int_{B_R}\frac{d\vec{\varepsilon}}{(2\pi\hbar)^\frac{3}{2} }\,e^{\frac{i}{\hbar}\vec\varepsilon\cdot\vec\pi}|\varepsilon_4| \left( \phi_+(\vec\varepsilon) - \phi_-(\vec\varepsilon)\right) \nonumber \\ & & + \int_{{\rm Fr}(B_R)}\frac{d\vec{\varepsilon}}{(2\pi\hbar)^\frac{3}{2} }\,e^{\frac{i}{\hbar}\vec\varepsilon\cdot\vec\pi} \phi_\bullet(\vec\varepsilon)
   \label{FourierExpbullet} \\ & & +\frac{1}{\hbar} \int_{{\rm Ext}(\bar B_R)}\frac{d\vec{\varepsilon}}{(2\pi\hbar)^\frac{3}{2} }\,e^{\frac{i}{\hbar}\vec\varepsilon\cdot\vec\pi} |i\varepsilon_4|\left( \varphi_+(\vec\varepsilon) -  \varphi_-(\vec\varepsilon) \right) \nonumber
\end{eqnarray}

Expanding in Fourier components the initial data, the functions $\phi_+(\vec\varepsilon),\phi_-(\vec\varepsilon), \phi_\circ(\vec\varepsilon), \phi_\bullet(\vec\varepsilon), \varphi_+(\vec\varepsilon)$ and 
$\varphi_-(\vec\varepsilon)$ can be solved for in terms of $\hat{\phicirc}(\vec\varepsilon)$ and $\hat{\phibullet}(\vec\varepsilon)$. Substituting these expressions in \eqref{FourierExp}, 
 a solution  $\phi(\vec\pi,\pi_4)$ of the IVP \eqref{IVP-Helmholtz-Appendix} is given by:
\begin{eqnarray}
 \phi(\vec\pi,\pi_4)&=&\int_{B_R}\frac{d\vec{\varepsilon}}{(2\pi\hbar)^\frac{3}{2} }\,e^{\frac{i}{\hbar}\vec\varepsilon\cdot\vec\pi} \left(
 \cos(|\varepsilon_4|\nu_4) \hat{\phicirc}(\vec\varepsilon) + \frac{\sin (|\varepsilon_4|\nu_4)}{|\varepsilon_4|} \hat{\phibullet}(\vec\varepsilon)\right)\nonumber \\
 & & + \int_{{\rm Fr}(B_R)}\frac{d\vec{\varepsilon}}{(2\pi\hbar)^\frac{3}{2} }\,e^{\frac{i}{\hbar}\vec\varepsilon\cdot\vec\pi} \left( \hat{\phicirc}(\vec\varepsilon) + \nu_4 \hat{\phibullet}(\vec\varepsilon)\right) \label{FourierExpIVP} \\ & & + \int_{{\rm Ext}(\bar B_R)}\frac{d\vec{\varepsilon}}{(2\pi\hbar)^\frac{3}{2} }\,e^{\frac{i}{\hbar}\vec\varepsilon\cdot\vec\pi} \left(\cosh(|i\varepsilon_4|\nu_4) \hat{\phicirc}(\vec\varepsilon) + \frac{\sinh (|i\varepsilon_4|\nu_4)}{|i\varepsilon_4|} \hat{\phibullet}(\vec\varepsilon) \right) \nonumber
\end{eqnarray}

\subsection{Uniqueness and well-posedness}

Let us discuss the Uniqueness and well-posedness of the IVP \eqref{IVP-Helmholtz-Appendix}. Consider two initial conditions $(\phicirc,\phibullet)$ and $(\phicirc{\!}',\phibullet{\!}')=(\phicirc,\phibullet)+(\delta\phicirc,\delta\phibullet)$, with $(\delta\!\!\phicirc,\delta\!\!\phibullet)$ a small perturbation. Denote by $\phi$ and $\phi'=\phi+\delta\phi$ the corresponding solutions of the 
IVP \eqref{IVP-Helmholtz-Appendix}. The we have that: 
\begin{eqnarray}
\delta\phi(\vec\pi,\pi_4)&=&\int_{B_R}\frac{d\vec{\varepsilon}}{(2\pi\hbar)^\frac{3}{2} }\,e^{\frac{i}{\hbar}\vec\varepsilon\cdot\vec\pi} \left(
 \cos(|\varepsilon_4|\nu_4) \widehat{\delta\!\!\phicirc}(\vec\varepsilon) + \frac{\sin (|\varepsilon_4|\nu_4)}{|\varepsilon_4|} \widehat{\delta\!\!\phibullet}(\vec\varepsilon)\right)\nonumber \\
 & & + \int_{{\rm Fr}(B_R)}\frac{d\vec{\varepsilon}}{(2\pi\hbar)^\frac{3}{2} }\,e^{\frac{i}{\hbar}\vec\varepsilon\cdot\vec\pi} \left( \widehat{\delta\!\!\phicirc}(\vec\varepsilon) + \nu_4 \widehat{\delta\!\!\phibullet}(\vec\varepsilon)\right) \label{UniquenessIVP} \\ & & + \int_{{\rm Ext}(\bar B_R)}\frac{d\vec{\varepsilon}}{(2\pi\hbar)^\frac{3}{2} }\,e^{\frac{i}{\hbar}\vec\varepsilon\cdot\vec\pi} \left(\cosh(|i\varepsilon_4|\nu_4) \widehat{\delta\!\!\phicirc}(\vec\varepsilon) + \frac{\sinh (|i\varepsilon_4|\nu_4)}{|i\varepsilon_4|} \widehat{\delta\!\!\phibullet}(\vec\varepsilon) \right) \nonumber
\end{eqnarray}

Suppose now that the two solutions are the same, i.e. $\delta\phi=0$. Evaluating \eqref{UniquenessIVP}  and its derivative with respect to $\nu_4$ at $\nu_4=0$, and equating them to zero we obtain that $\delta\!\!\phicirc=0$ and $\delta\!\!\phibarbullet=0$, and therefore the solution is unique.

To study well-posedness, from \eqref{UniquenessIVP} we see that $|\delta\phi(\vec\pi,\pi_4)|$ is bounded in $\pi_4$ iff supp$(\widehat{\delta\!\!\phicirc})\subset \bar B_R$ and supp$(\widehat{\delta\!\!\phibullet})\subset B_R$ (i.e. iff $ \widehat{\delta\!\!\phicirc}=0$ for $\|\vec\varepsilon\|> R$ and $ \widehat{\delta\!\!\phibullet}=0$ for $\|\vec\varepsilon\|\geq R$).

\subsection{Oscillatory solutions}

Solutions to the IVP \eqref{IVP-Helmholtz-Appendix} satisfying that supp$(\hat{\phicirc})\subset \bar B_R$ and supp$(\hat{\phibullet})\subset B_R$  are usually known as \textit{oscillatory solutions}  (since the behaviour with respect to $\nu_4$ is oscillatory or constant) \cite{Steinberg-Wolf}, and for them then the IVP is well-posed. 

More precisely, denote by ${\cal V}_m$  the linear subspace of solutions to the IVP  \eqref{IVP-Helmholtz-Appendix}  for which 
supp$(\hat{\phicirc})\subset \bar B_R$,  and  supp$(\hat{\phibullet})\subset B_R$. Then 
${\cal V}_m$ is the maximal subspace of solutions   for which the IVP \eqref{IVP-Helmholtz-Appendix} is well-posed, and it is known as \textit{oscillatory solutions}.

To construct the unitary and irreducible representation in momentum space, we shall restrict to  ${\cal V}_m$, since otherwise the representation would not be unitary. Also, the subspace ${\cal V}_m$ is invariant under the action of the group $\tilde\Sigma(SU(2))$ (and the Euclidean group $E(4)$), as can be easily checked.

Note that, when restricted to oscillatory solutions, Eq. \eqref{FourierExp}  (restricted to $B_R$) coincides with Eq. \eqref{FourierUp} of the Generalized Fourier Transform (given in Sec. \ref{Fourier}) if we make the identifications:
\begin{equation}
  \Psi^\pm(\vec\varepsilon\,) = |\varepsilon_4|_+  \phi_\pm(\vec\varepsilon) 
\end{equation}

\section{Digression on Hilbert spaces, Dirac's bra-ket formulation and resolution of the identity operators}
\label{DiracNotation}

Let us review Dirac's bra-ket formulation, which is well known in quantum mechanics, but that has to be  applied with care in the cases we are discussing in this paper. We shall also discuss the 
definition of resolution operators and their properties.

In Dirac's bra-ket formulation, given a Hilbert space ${\cal H}$ with scalar product $\langle\cdot,\cdot\rangle$, we can denote elements of the antidual space (i.e. antilinear functionals on ${\cal H}$) $\Phi\in {\cal H}^\times$ as a \textit{ket} $|\Phi\rangle$. 
Elements of the dual space (i.e. linear functionals on ${\cal H}$) $\Psi\in {\cal H}^*$ are denoted as bras $\langle \Psi|$. 

Using Riesz representation theorem\footnote{Note that we are using physicists convention for the complex scalar product, i.e. anti-linear in the first entry.} \cite{RieszRepresentationTheorem}, for any ket $|\Phi\rangle$ there exists $\phi\in {\cal H}$ such that $\langle \Phi|(\chi)=\langle \chi,\phi\rangle\,,\forall \chi\in {\cal H}$. Also, for any bra $\langle \Psi|$ there exists 
$\psi\in {\cal H}$ such that $\langle \Psi|(\chi)=\langle \psi,\chi\rangle\,,\forall \chi\in {\cal H}$. By abuse of notation we shall write  $\langle \psi|$ for the  bra  $\langle \Psi|$ and $|\phi\rangle$ for the ket $|\Phi\rangle$.

The antidual space ${\cal H}^\times$ can be identified with ${\cal H}$, in such a way that we can define the action of a bra $\langle \Psi|$ on a ket $|\Phi\rangle$ as:
\begin{equation}
 \langle \Psi| (|\Phi\rangle)\equiv \langle \Psi|\Phi\rangle=\langle \psi,\phi\rangle
\end{equation}
Note that $\langle \psi|\phi\rangle = \langle \psi,\phi\rangle
= \langle \phi,\psi\rangle^*=\langle \phi|\psi\rangle^*$.

However, in general $\langle \psi| \neq |\psi \rangle^\dag$, i.e. the bra is not
the transpose conjugate of the ket (seen as a column vector). It is true in the standard Hilbert space $L^2(\mathbb{R},dx)$ (or the general case
$L^2(\mathbb{R}^n,d^n\vec x)$), but it is
not true in any Hilbert space of the form $L^2(\mathbb{R},\mu(x)dx)$, with $\mu>0$. If we denote by $\hat\mu$ the positive
multiplicative operator $\hat\mu \phi(x)=\mu( x) \phi( x)$, then $\langle \psi| = (\hat\mu\,|\psi \rangle)^\dag= |\psi \rangle^\dag\hat\mu$.

For the more general case of the Hilbert space $L^2_{\hat{K}}(\mathbb{R})$ (or its higher-dimensional generalizations), the scalar product is given by:
\begin{equation}
 \langle \psi ,\phi\rangle_{\hat{K}}=\langle \psi ,\hat{K}\phi\rangle
\end{equation}
where $\langle \psi ,\phi\rangle$ is the scalar product in $L^2(\mathbb{R},dx)$ and $\hat{K}$ is a positive definite self-adjoint operator on 
$L^2(\mathbb{R},dx)$. In general $\hat{K}$ will be an integral operator, with (possibly distributional) kernel $k(x,x')$:
\begin{equation}
 \hat{K}\phi(x)=\int_\mathbb{R} dx k(x,x')\phi(x')
\end{equation}
Thus the scalar product can be written as a double integral:
\begin{equation}
 \langle \psi ,\phi\rangle_{\hat{K}}=\int_\mathbb{R} dx'\int_\mathbb{R} dx \psi(x)^*k(x,x')\phi(x')
\end{equation}

The previous case of $L^2(\mathbb{R},\mu(x)dx)$  is recovered when $k(x,x')=\mu(x)\delta(x-x')$. In $L^2_{\hat{K}}(\mathbb{R})$ we have that $\langle \psi| = (\hat K\,|\psi \rangle)^\dag= |\psi \rangle^\dag\hat K$. In general, the adjoint $\hat{A}^\dag_{\hat{K}}$ of an operator $\hat A$ on $L^2_{\hat{K}}(\mathbb{R})$ is (see, for instance, \cite{Ali}):
\begin{equation}
 \hat{A}^\dag_{\hat{K}}=\hat{K}^{-1}\hat{A}^\dag\hat{K}
 \label{kadjunto}
\end{equation}
where $\hat{A}^\dag$ is the adjoint in $L^2(\mathbb{R},dx)$.

A particularly interesting case is when $k(x,x')=k(x-x')$. Then the operator $\hat{K}$ is a \textit{convolution} operator, i.e $\hat{K}\phi(x) = k*\phi (x)$. Convolution operators are diagonalized by the Fourier transform, in such a 
way that 
\begin{equation}
 {\cal F}(\hat{K}\phi)(p)= {\cal F}(k)(p){\cal F}(\phi)(p)
\end{equation}

In other words, $\hat{K}$ is a multiplier operator, with multiplier $m_k(p)={\cal F}(k)(p)$ in Fourier domain. The convolution operator $\hat{K}$ is positive definite if and only if $m_k(p)>0$ for all $p\in\mathbb{R}$.

Let us  denote by $|\psi\rangle\langle\phi|$ the linear operator on ${\cal H}$ defined as:
\begin{equation}
 |\psi\rangle\langle\phi|(\phi')= |\psi\rangle\langle\phi|(|\phi'\rangle)\equiv |\psi\rangle\langle\phi|\phi'\rangle= \langle\phi,\phi'\rangle |\psi\rangle =  \langle\phi,\phi'\rangle\, \psi
\end{equation}

Dirac's formalism is valid for more general settings than Hilbert spaces, like rigged Hilbert spaces \cite{BohmGadella} or even in Banach spaces. In this case, given a Hilbert space ${\cal H}$, choose a suitable dense
subspace ${\cal S}\subset {\cal H}$ of \textit{test functions}. Then we have:
\begin{equation}
 {\cal S}\subset {\cal H}\subset {\cal S}^*\,,\qquad {\cal S}\subset {\cal H}\subset {\cal S}^\times
\end{equation}
Kets are elements ${\cal S}^\times$ and bras are elements of ${\cal S}^*$. By the generalized Riesz representation theorem, to a ket $\Phi\in {\cal S}^\times$ we can associate a distribution $\phi$ (non-normalizable, in general)
such that $\Phi(\chi)=\langle \chi,\phi\rangle\,,\forall \chi\in {\cal S}$. In the same way, to a bra $\Psi\in {\cal S}^*$ we can associate a distribution $\psi$ (non-normalizable, in general)
such that $\Psi(\chi)=\langle \psi,\chi\rangle\,,\forall \chi\in {\cal S}$.

In particular, in the standard case of ${\cal H}_x=L^2(\mathbb{R}^3,d\vec x)$, we define $|\vec{x}\,'\rangle$ as the ket associated with the distribution $\phi_{{\vec x}\,'}(\vec x)=\delta^{(3)}(\vec x-\vec{x}\,')$. The states $|\vec{x}\rangle$ are (generalized)
eigenstates of position operator $\hat{\vec x}\phi(x)=\vec x\phi(x)$, and constitute a complete family of states expanding ${\cal H}_x$, in the sense that:
\begin{equation}
 I_{{\cal H}_x} = \int_{\mathbb{R}^3}d\vec x |\vec x\rangle \langle \vec x|
 \label{ResolutionIdentityx}
\end{equation}
where $I_{{\cal H}_x}$ is the identity operator in ${\cal H}_x$, and the convergence of the integral should be understood in the weak sense \cite{AliAntoineGazeau}. The measure $d\vec x$  in the previous integral is the spectral measure of the operator  $\hat{\vec x}$. 
Equation \eqref{ResolutionIdentityx} is known as a \textit{resolution of the identity} operator.
Note that a  wavefunction in configuration space $\psi(\vec{x})\in {\cal S}_x$, with ${\cal S}_x$ a suitable subspace of test functions, can be written as $\psi(\vec x)=\langle\vec x|\psi\rangle=\langle \phi_{{\vec x}},\psi\rangle$,

\section{Some identities for Bessel related functions}
\label{AppendixBessel}

Let us enumerate some identities related to Bessel functions of the type appearing in the quantization in momentum space.
Denote by $B^{(d)}=\{x\in\mathbb{R}^d\,/\,\|\vec x\|< 1\}$ the unit ball in $d$ dimensions, and 
\begin{equation}
 (x)_+=\left\{ \begin{array}{lcl}
                x & {\rm if} & x> 0 \\
                0 & {\rm if} & x\leq 0
               \end{array}\right.
\end{equation}
the continuous function which is usually known as the \text{ramp} function $R(x)$, and that can also be written as $R(x)=\frac{x+|x|}{2}=xH(x)$, with $H(x)$ is Heaviside function. 
Define the functions $(x)_+^\alpha$ as
\begin{equation}
 (x)_+^\alpha=\left\{ \begin{array}{lcl}
                x^\alpha & {\rm if} & x> 0 \\
                0 & {\rm if} & x\leq 0
               \end{array}\right.
\end{equation}
for $\alpha\in\mathbb{R}$. Note that  $(x)_+^0\equiv\chi_{\mathbb{R}^+}(x)$, where $\chi_A$ is the indicator function of the subset $A\subset \mathbb{R}$.

Then  
\begin{equation}
 \int_{\mathbb{R}^d}\frac{d\vec{x}}{(2\pi)^\frac{d}{2} }\,(1-\|\vec x\|^2)^\alpha_+ e^{i \vec x\cdot \vec p}  =  \int_{B^{(d)}}\frac{d\vec{x}}{(2\pi)^\frac{d}{2} }\, (1-\|\vec x\|^2)^\alpha e^{i \vec x\cdot \vec p} = \frac{1}{N_{\alpha}}   \frac{J_{\frac{d}{2}+\alpha}(\|\vec p\|)}{\|\vec p\|^{\frac{d}{2}+\alpha}}\,,\qquad\alpha>-1
 \label{besselar}
\end{equation}

\begin{equation}
 \int_{\mathbb{R}^d} \frac{d\vec{p}}{(2\pi)^\frac{d}{2} }\,
\frac{J_\alpha(\|\vec p\|)}{\|\vec p\|^\alpha}
e^{-i \vec x \cdot \vec p } = N_{\alpha-\frac{d}{2}} (1-\|\vec x\|^2)^{\alpha-\frac{d}{2}}_+ \,\,,\qquad \alpha>\frac{d}{2}-1
\label{desbeselar}
\end{equation}
where $N_\alpha=\frac{1}{2^{\alpha}\Gamma(\alpha+1)}$.
Let us denote by $k_\alpha(\vec p)= \frac{J_\alpha(\|\vec p\|)}{\|\vec p\|^\alpha}$. Then

\begin{equation}
 \hat{K}_\alpha\phi(\vec p)=\int_{\mathbb{R}^d} d\vec p\, k_\alpha(\vec p - \vec p\,')\phi(\vec p\,')\,,\qquad  \alpha> \frac{d}{2}-1
\end{equation}
is a non-negative operator, since $m_{k_\alpha}(\vec x)= N_{\alpha-\frac{d}{2}}  (1-\|\vec x\|^2)^{\alpha-\frac{d}{2}}_+$ is a non-negative multiplier. These operators are known as Bochner-Riesz means  or Bochner-Riesz integral kernels \cite{Bochner-Riesz-mean}.

The convolution kernels satisfy the \textit{reproducing} property:
\begin{equation}
 \int_{\mathbb{R}^d} d\vec p\,'' \, k_\alpha(\vec p-\vec p\,'') k_\beta(\vec p\,''-\vec p\,') = \frac{N_{\alpha-\frac{d}{2}} N_{\beta-\frac{d}{2}}}{N_{\alpha+\beta-d}}k_{\alpha+\beta-\frac{d}{2}}(\vec p-\vec p\,')\,,\qquad \alpha,\beta> \frac{d}{2}-1\,,\quad \alpha+\beta> d-1 
  \label{reproducing}
\end{equation}
from which the rule for the product of two operators can be deduced:
\begin{equation}
 \hat{K}_\alpha \hat{K}_\beta = \frac{N_{\alpha-\frac{d}{2}} N_{\beta-\frac{d}{2}}}{N_{\alpha+\beta-d}} \hat{K}_{\alpha+\beta-\frac{d}{2}}\,,\qquad  \alpha+\beta> d-1 
\end{equation}

\section{On the Hilbert spaces $L^2_{\hat{K}_\alpha}(\mathbb{R}^d)$}
\label{SpacesLk}

Denote by $PW_1(\mathbb{R}^d)\equiv PW$ the Paley-Wiener \cite{PaleyWiener}  subspace of $L^2(\mathbb{R}^d)$ made of square-integrable functions on $\mathbb{R}^d$ whose Fourier transform is supported on $B^{(d)}$ (i.e. \textit{bandlimited functions} on $\mathbb{R}^d$). Denote by $\widetilde{PW}$ the  Paley-Wiener-Schwarz \cite{Hormander}  Hilbert space of functions on $\mathbb{R}^d$ whose
Fourier transform is a distribution supported on $B^{(d)}$.

Then the  operators $\hat{K}_\alpha$ turn out to be positive definite on $\widetilde{PW}$, and therefore they define the family of (reproducing kernel) Hilbert spaces $L^2_{\hat{K}_\alpha}(\mathbb{R}^d){\subset} \widetilde{PW}$, 
with convolutions kernels $k_\alpha$, according to Appendix \ref{DiracNotation}.

Note that the multiplier associated with $\hat{K}_{\frac{d}{2}}$ is precissely the indicator function $(x)_+^0$. Therefore, $\hat{K}_{\frac{d}{2}}$ is the projector onto $\widetilde{PW}$, i.e. 
$\hat{K}_{\frac{d}{2}}f\in \widetilde{PW}$ for any $f\in\widetilde{PW}$ and $\hat{K}_{\frac{d}{2}}^2f=\hat{K}_{\frac{d}{2}}f$. In fact, $\hat{K}_{\frac{d}{2}}$  plays the role of the Identity operator on $\widetilde{PW}$. Note that for $d=1$, the convolution kernel is $k_{1/2}(p)={\rm sinc}(p)$, thus $k_{d/2}$ generalizes the well-known sinc function of signal analysis and sampling theory for higher dimensions.

%
%
%

Using the expression for the multiplier of $\hat{K}_{\alpha}$, it is easy to obtain the  spectrum of these operators on $\widetilde{PW}$:
\begin{equation}
  \text{spec}(\hat{K}_{\alpha})=\left\{ 
                                       \begin{array}{cc}
                                        [N_{\alpha-\frac{d}{2}},\infty) & \alpha<\frac{d}{2} \\
                                        1 & \alpha=\frac{d}{2}\\
                                        (0,N_{\alpha-\frac{d}{2}}] & \alpha>\frac{d}{2}
                                       \end{array}
                                       \right.
\end{equation}

With this, it is easy to obtain the following inclusions \cite{Ali}:
\begin{equation}
\begin{array}{cc}
 L^2_{\hat{K}_\alpha}(\mathbb{R}^d) \subset PW & \frac{d}{2}-1 <\alpha < \frac{d}{2} \\
 L^2_{\hat{K}_\alpha}(\mathbb{R}^d) = PW & \alpha = \frac{d}{2} \\
 PW \subset L^2_{\hat{K}_\alpha}(\mathbb{R}^d) & \alpha > \frac{d}{2}
\end{array}
\end{equation}

Also, for $\frac{d}{2}-1 <\alpha < \frac{d}{2}$, we have:
\begin{equation}
 L^2_{\hat{K}_\alpha}(\mathbb{R}^d) \subset PW \subset 
 L^2_{\hat{K}_{d-\alpha}}(\mathbb{R}^d)
\end{equation}

Therefore the Hilbert spaces $L^2_{\hat{K}_\alpha}(\mathbb{R}^d)$, $PW$ and $L^2_{\hat{K}_{d-\alpha}}(\mathbb{R}^d)$ define a Gelfand triple of 
(rigged) Hilbert spaces \cite{Ali}.

\section{Normalizing constants for stationary states in momentum space}
\label{Mmomentos}

In this Appendix we compute the explicit expression of the normalizing constants $M_{nlm}$ appearing
in equation \eqref{wavefunctionsmom}. Since $\pi_4$ is singularized in 
\eqref{ProductoEscalarMomentosarriba}, we need to perform a change of variables in 
\eqref{wavefunctionsmom} from hyperspherical to  hypercylindrical ones 
$(\pi_\rho, \pi_\theta, \pi_\phi, \pi_4)$, 
\begin{align}
\pi_\rho &=\,\pi_r \sin{\pi_\chi}\nn\\
\pi_4 &=\, \pi_r \cos{\pi_\chi}\,.
\end{align}
We then evaluate $\phi_{nlm}(\pi_\rho,\pi_\theta,\pi_\phi,\pi_4 = 0)$ and 
$\frac{\partial\,}{\partial \pi_4}\phi_{nlm}(\pi_\rho,\pi_\theta,\pi_\phi,\pi_4 = 0)$ 
to obtain the input data for the energy eigenstates $\phicirc_{nlm}$ and 
$\phibullet_{nlm}$: 
\begin{align}
\phicirc_{nlm}(\vec{\pi}) &=M_{nlm}\frac{1}{\pi_\rho}C^{(l+1)}_{n-l}(0)Y_{lm}(\pi_\theta,\pi_\phi) \big(A \,J_{n+1}(\kappa \pi_\rho)+B \,Y_{n+1}(\kappa \pi_\rho)\big)\nn\\
\phibullet_{nlm}(\vec{\pi}) &=M_{nlm}\frac{2(l+1)}{\pi_\rho^2}C^{(l+2)}_{n-l-1}(0)Y_{lm}(\pi_\theta,\pi_\phi) \big(A \,J_{n+1}(\kappa \pi_\rho)+B \,Y_{n+1}(\kappa \pi_\rho)\big)\,,
\label{wavefunctionsmomcomp} 
\end{align}
where $\vec\pi$ is written in usual spherical coordinates 
$(\pi_\rho,\pi_\theta,\pi_\phi)$. With that, we can now use
\eqref{ProductoEscalarMomentos} in order to establish orthonormalizability, the 
quantization of $n$, $l$ and $m$ and that only $B=0$ is allowed.  

Input data \eqref{wavefunctionsmomcomp} for eigenstates of the Hamiltonian in momentum 
space \eqref{wavefunctionsmom} can be used to establish orthogonality and the 
normalizing constants $M_{nlm}$. We want to compute
\[
	\langle \phi_{n'l'm'},\phi_{nlm}\rangle_m
	=  C
\int_{\mathbb{R}^3}d\vec{\pi}\int_{\mathbb{R}^3}d\vec{\pi}{}' \left( \phicirc(\vec\pi')^*_{n'l'm'}  
\kappa^2 k_2(\kappa \|\vec\pi-\vec\pi'\|) \phicirc_{nlm}(\vec\pi) 
+ \phibullet(\vec\pi')^*_{n'l'm'} 
k_1(\kappa \|\vec\pi-\vec\pi'\|) \phibullet_{nlm}(\vec\pi) \right)\,.
\]
Using spherical momenta $(\pi_\rho,\pi_\theta,\pi_\phi)$, the integrals in the 
angular part of $\vec\pi$ can be evaluated by taking into account 
formulas \eqref{besselar} in Appendix \ref{AppendixBessel}, 
together with the 
well-known expansion of plane waves in spherical waves 
$e^{ \frac{i}{\hbar}\vec \epsilon \cdot \vec \pi} = 
4\pi \sum_{k=0}^{\infty}\sum_{s=-k}^{+k}i^k 
j_k(\frac{1}{\hbar}\epsilon_\rho \pi_\rho)Y_{ks}(\epsilon_\theta,\epsilon_\phi)
Y_{ks}^*(\pi_\theta,\pi_\phi)$, where $j_k(x)$ are the spherical Bessel functions. 
The angular integration in $\vec\pi'$ can then be computed by using the orthogonality 
of spherical harmonics appearing in $\phicirc_{nlm}(\vec\pi)$ and 
$\phibullet_{nlm}(\vec\pi)$ to give: 
\begin{multline*}
	\langle \phi_{n'l'm'},\phi_{nlm}\rangle_m = 
	\\ 
 =  \delta_{ll'}\delta_{mm'} M_{n'lm}M_{nlm} C \Bigg\{
  \frac{8}{\kappa^2\hbar^4}C^{l+1}_{n'-l}(0)C^{l+1}_{n-l}(0)
 \int_0^\infty d\pi_\rho' \pi_\rho' 
 \Big(A' \,J_{n'+1}(\kappa \pi_\rho')+B' \,Y_{n'+1}(\kappa \pi_\rho')\Big)
 \times
 \\
 \int_0^R d\epsilon_\rho \epsilon_\rho^2 \sqrt{R^2-\epsilon_r^2} j_l(\frac{1}{\hbar}\epsilon_\rho \pi_\rho') 
 \int_0^\infty d\pi_\rho \pi_\rho j_l(\frac{1}{\hbar}\epsilon_\rho \pi_\rho) 
 \Big(A \,J_{n+1}(\kappa \pi_\rho)+B \,Y_{n+1}(\kappa \pi_\rho)\Big) 
 \\
 +
  \frac{32 (l+1)^2}{\hbar^2\kappa^2}C^{l+2}_{n'-l-1}(0)C^{l+2}_{n-l-1}(0)
 \int_0^\infty d\pi_\rho'  
 \Big(A' \,J_{n'+1}(\kappa \pi_\rho')+B' \,Y_{n'+1}(\kappa \pi_\rho')\Big)\times
 \\
 \int_0^R \frac{d\epsilon_\rho \epsilon_\rho^2}{\sqrt{R^2-\epsilon_r^2}}   j_l(\frac{1}{\hbar}\epsilon_\rho \pi_\rho') 
 \int_0^\infty d\pi_\rho  j_l(\frac{1}{\hbar}\epsilon_\rho \pi_\rho) 
 \Big(A \,J_{n+1}(\kappa \pi_\rho)+B \,Y_{n+1}(\kappa \pi_\rho)\Big)
 \Bigg\}\,.
\end{multline*}

$A$ and $A'$ can be arbitrarily set to $1$. For $B'=B=0$ and $n$ and $n'$ integers, 
$n'\geq l\geq 0$, $n\geq l\geq 0$, the integrals appearing above converge, giving: 
\begin{align*}
	\int_0^\infty d\pi_\rho \pi_\rho j_l(\frac{1}{\hbar}\epsilon_\rho \pi_\rho) 
 J_{n+1}(\kappa \pi_\rho) &= \frac{K_{nl}^\circ}{\kappa^2}
 \frac{\left(\frac{\epsilon_r}{R}\right)^l}{\sqrt{1-\frac{\epsilon_r^2}{R^2}}} C_{n-l}^{l+1}\Big(\sqrt{1-\frac{\epsilon_r^2}{R^2}}\Big)\,,
 \\
 \int_0^\infty d\pi_\rho j_l(\frac{1}{\hbar}\epsilon_\rho \pi_\rho) 
 J_{n+1}(\kappa \pi_\rho) &= \frac{K_{nl}^\bullet}{\kappa}
 \frac{\left(\frac{\epsilon_r}{R}\right)^l}{\sqrt{1-\frac{\epsilon_r^2}{R^2}}} C_{n-l}^{l+1}\Big(\sqrt{1-\frac{\epsilon_r^2}{R^2}}\Big)\,,
\end{align*}
where
\begin{align*}
	K_{nl}^\circ &= \frac{-\pi^{\frac{3}{2}} 2^{l-n-1}\Gamma(l+1)e^{i\frac{\pi}{2}(l-n)}\Big(e^{in\pi}\csc{(\frac{\pi}{2}(l+n))+i\sec{(\frac{\pi}{2}(l-n)})\Big)}}{\Gamma(l-n)\Gamma(\frac{1}{2}(n-l+1))\Gamma(\frac{1}{2}(n+l+2))}\,,
 \\
 K_{nl}^\bullet &=\frac{\pi^{\frac{3}{2}} 2^{l-n-2}\Gamma(l+1)e^{i\frac{\pi}{2}(l-n)}\Big(\csc{(\frac{\pi}{2}(l-n))+ie^{in\pi}\sec{(\frac{\pi}{2}(l+n)})\Big)}}{\Gamma(l-n)\Gamma(\frac{1}{2}(n-l+2))\Gamma(\frac{1}{2}(n+l+3))}
 \,.
\end{align*}
For integers $n$ and $l$, $n\geq l\geq 0$,  $K_{nl}^\circ$ and $K_{nl}^\bullet$ prove to be 
real numbers. 

We distinguish two cases: 
\begin{itemize}
	\item $n-l$ even. In this case, $C^{l+2}_{n-l-1}(0)=0$. Making the change 
	      $\epsilon_4 = +\sqrt{R^2-\epsilon_r^2}$, and using $C=\frac{\hbar\kappa^2}{4\pi}$ and that $C^{l+1}_{n-l}(x)$
	      is even in $x$ if $n-l$ is even, we get: 
	\begin{multline*}
		\langle \phi_{n'l'm'},\phi_{nlm}\rangle_m = \delta_{ll'}\delta_{mm'} M_{n'lm}M_{nlm} \frac{\hbar}{\pi}C^{l+1}_{n'-l}(0)C^{l+1}_{n-l}(0) K_{n'l}^\circ K_{nl}^\circ \int_{-R}^{R} \frac{d \epsilon_4}{R} \big(1-\frac{\epsilon_4^2}{R^2} \big)^{l+\frac{1}{2}}C^{l+1}_{n'-l}(\frac{\epsilon_4}{R})C^{l+1}_{n-l}(\frac{\epsilon_4}{R})\\
		= \delta_{ll'}\delta_{mm'} \delta_{nn'}M_{nlm}^2 
		(C^{l+1}_{n-l}(0))^2 (K_{nl}^\circ)^2
		\frac{\hbar \Gamma(n+l+2)}{2^{2l+1} (n+1)\Gamma(l+1)^2} \,, 
	\end{multline*}
	 from which $M_{nlm}$ can be computed up to a sign: 
	 \[
	 M_{nlm}= \frac{i 2^{l+1} l! e^{i\frac{\pi}{2}(n-l)}\csc{(\pi(l-n))}}{1 + i e^{i\pi n}\cos{(\frac{\pi}{2}(l-n))}\csc{(\frac{\pi}{2}(n+l))}}\sqrt{\frac{2(n+1)}{\hbar(n+l+1)!}}\,.
	 \]
	 The first values are given by $M_{00m}=\frac{\sqrt{2}}{\sqrt{\hbar}}$, $M_{20m}=-\frac{1}{\sqrt{\hbar}}$, $M_{11m}=\frac{2\sqrt{2}}{\sqrt{3\hbar}}$, \ldots
	\item $n-l$ odd. Now $C^{l+1}_{n-l}(0)=0$. Following similar steps we find: 
	\[
	\langle \phi_{n'l'm'},\phi_{nlm}\rangle_m = 
	\delta_{ll'}\delta_{mm'} \delta_{nn'}M_{nlm}^2 
		(C^{l+1}_{n-l}(0))^2 (K_{nl}^\bullet)^2
		\frac{ (l+1)^2 \hbar \Gamma(n+l+2)}{2^{2l-1} (n+1)\Gamma(l+1)^2}\,, 
	\]
	and 
	\[
	 M_{nlm}= -\frac{ 2^{l+1} l! e^{i\frac{\pi}{2}(n-l)}\csc{(\pi(l-n))}}{1 + i e^{i\pi n}\sin{(\frac{\pi}{2}(l-n))}\sec{(\frac{\pi}{2}(n+l))}}\sqrt{\frac{2(n+1)}{ \hbar (n+l+1)!}}\,.
	 \]
	 The first values are given by $M_{10m}=\frac{\sqrt{2}}{\sqrt{\hbar}}$, $M_{30m}=-\frac{1}{\sqrt{3\hbar}}$, $M_{32m}=\frac{1}{\sqrt{\hbar}}$, \ldots
\end{itemize}


\bigskip





\begin{thebibliography}{99}

\bibitem{SU(2)} V. Aldaya, J. Guerrero, F.F. L\'opez-Ruiz and F. Coss\'{i}o, \textit{SU(2) particle sigma model: the role of
contact symmetries in global quantization}, J. Phys. {\bf A49}, (2016) 505201.


\bibitem{Gadella} M. Gadella, L.M. Nieto, J. Negro, G.P. Pronko, M. Santander, \textit{Spectrum Generating Algebras for the free
motion in $S^3$},  J. Math. Phys. \textbf{52}, 063509 (2011)

\bibitem{Santander1} J.F. Cari\~nena, M. Ra\~nada, M. Santander, \textit{The quantum free particle on spherical and hyperbolic spaces: A curvature dependent approach}, J. Math. Phys. \textbf{52}, 072104 (2011)

\bibitem{Santander2} J.F. Cari\~nena, M. Ra\~nada, M. Santander, \textit{The quantum free particle on spherical and hyperbolic spaces: A curvature dependent approach. II }, J. Math. Phys. \textbf{53}, 102109 (2012)

\bibitem{Liu} Q.H. Liu, L.H. Tang, and D.M. Xun, \textit{Geometric momentum: The proper momentum for a free particle on a two-dimensional sphere}, Phys. Rev. A \textbf{84}, 042101 (2011)

\bibitem{RMP} Aldaya, Calixto, Guerrero, L\'opez-Ruiz, \textit{Group Quantization of non-linear sigma models: particle on $S^2$ revisited}, Rep. Math. Phys. \textbf{64}, 49-58 (2009)



\bibitem{Folland} G.B. Folland, \textit{A Course in Abstract Harmonic Analysis}, CRC Press, Boca Raton, FL (1995)

\bibitem{GeneralPhaseSpaces} V. V Albert, S. Pascazio and M. H. Devoret, \textit{General phase spaces: from discrete
variables to rotor and continuum limits}, J. Phys. A \textbf {50} 504002 (2017)

\bibitem{Sherman} T.O. Sherman, \textit{Fourier Analysis on the Sphere}, 
Trans. Am. Math. Soc. \textbf{209}, 1-31 (1975)


\bibitem{Volobuyev} I.P. Volobuyev, \textit{Plane waves on a sphere and some applications}, Teor. Mat. Fiz. \textbf{45}, 421 (1980)



\bibitem{WolfWignerFunction-Sphere} M.A. Alonso, G.S. Pogosyan and K. B. Wolf, \textit{Wigner functions for curved spaces. II. On spheres},  J. Math. Phys. \textbf{44}, 1472-1489 (2003)

\bibitem{BernardoOptica} K. B. Wolf, \textit{Elements of Euclidean optics}, in \textit{Lie Methods in Optics}, Lecture Notes in Physics,
Springer-Verlag, Heidelberg,  115 (1989)



 \bibitem{23} V. Aldaya  and J.A. de Azc\'arraga,  J. Math. Phys. \textbf{23}, 1297 (1982)

\bibitem{Aldaya-review-GAQ} V. Aldaya, J. Guerrero, \textit{Lie Groups Representations and Quantization}, Rep. Math.
Phys. \textbf{47 }(2001) 213.

\bibitem{Souriau} J.M. Souriau, \textit{Structure of Dynamical
Systems}, Birkh\"auser (1997).

\bibitem{Kostant} B. Kostant, \textit{Quantization and unitary representations, }\textcolor{black}{Lec.
Notes in Math. }\textbf{170}, Springer (1970).

\bibitem{Kirillov} A.A. Kirillov, \textit{Elements of the Theory of Representacions},
Springer Verlag (1976).

\bibitem{Quasi-invariant-measure} J. Guerrero, V. Aldaya, \textit{J. Math. Phys.} \textbf{41} (2000) 6747.


\bibitem{HOPolarization1} V. Aldaya, J. Guerrero and G. Marmo, Int. J. Mod. Phys. A \textbf{12}, 3-11 (1997)

\bibitem{HOPolarization2}  V. Aldaya and J. Guerrero, \textit{Lie Group Representations
and Quantization},  Rep. Math. Phys. \textbf{47} (2001)

\bibitem{Payne} L. E. Payne, \textit{Improperly Posed Problems in Partial Differential Equations}, SIAM, Philadelphia (1975)

\bibitem{Steinberg-Wolf} S.Steinberg and K.B. Wolf, \textit{Invariant inner products on spaces of
solutions of the Klein-Gordon and Helmholtz equations}, J. Math. Phys. \textbf{22}, 1660-1663 (1981)

\bibitem{Gustavo} Gustavo Garrig\'os, private communication.


\bibitem{PositionOperator} V. Aldaya, J. Bisquert, J. Guerrero and J. Navarro-Salas, J. Phys A \textbf{26}, 5375-5390 (1993)


\bibitem{Salpeter} E. E. Salpeter, \textit{Mass Corrections to the Fine Structure of Hydrogen-Like Atom},  Phys. Rev. \textbf{87} 328--343 (1952)


\bibitem{Salpeter2} B. Rosenstein \& M. Usher, \textit{Explicit illustration of causality violation: Noncausal relativistic wave-packet evolution}, Phys.  Rev. \textbf {D 36}, 2381--2384 (1987)

\bibitem{KowalskiSalpeter} K. Kowalski, J. Rembieli\'nski and J.-P. Gazeau, \textit{On the coherent states for a relativistic scalar particle}, Ann. Phys. \textbf{399}, 204-223 (2018)

\bibitem{BernardoManko}   V. I. Man'ko, K. B. Wolf, \textit{The map between Heisenberg-Weyl and Euclidean optics is comatic}, in \textit{Lie Methods in Optics},  p. 163. Lecture Notes in Physics,
Springer-Verlag, Heidelberg  (1989)


\bibitem{AliAntoineGazeau}   S.T. Ali, J.P. Antoine, J.P. Gazeau, \textit{Coherent States, Wavelets, and Their Generalizations}, Springer, New York (2014)



\bibitem{deBievre}  S. de Bi\`evre,  Coherent states over symplectic homogenous spaces, J Math Phys \textbf{30}  (1989) 1401-1407




\bibitem{deBievreGonzalez} S. De Bi\`evre and J. A. Gonz\'alez, in \textit{Quantization and Coherent States Methods}, Proc. XIth Workshop on
Geometric Methods in Physics, Bia\l owieza, Poland, 1992, eds. S. Twareque Ali, I. M. Mladenov, and A. Odzijewicz, World Scientific, Singapore, 1993, p. 152.

\bibitem{Kowalski} K. Kowalski, J. Rembieli\'nski and L. C. Papaloucas, \textit{Coherent states for a quantum particle on a circle}, J. Phys. A: Math. Gen. \textbf{29}, 4149 (1996)

\bibitem{MOlmoCS} J.A.  Gonz\'alez and M.A. del Olmo,  Coherent states on the circle and quantization,  
J. Phys. A \textbf{31} (1998) 8841-8857


\bibitem{Kastrup} H. A. Kastrup, Quantization of the canonically conjugate pair angle and orbital angular momentum
Phys. Rev. A \textbf{73} (2006) 052104

\bibitem{GazeauPLA} P.L. Garc\'\i a de Le\'on and J.P. Gazeau, Coherent state quantization and phase operator, 
Physics Letters \textbf{A361} (2007) 301-304



\bibitem{IshamKlauder} C.J. Isham and J.R. Klauder, Coherent states for n-dimensional Euclidean groups E(n) and their
application, J Math Phys \textbf{32} (1991) 607-620

\bibitem{GazeauCS}  R. Fresneda, J. P. Gazeau and D. Noguera, Quantum localisation on the circle
J. Math. Phys. \textbf{59} (2018) 052105 


\bibitem{WignerFunctionEuclieo} L.M. Nieto, N.M. Atakishiyev, S.M. Chumakov  and K.W. Wolf   \textit{Wigner distribution function
for Euclidean systems}, J. Phys. A \textbf{31} 3875--3895 (1998)
 
 
\bibitem{WignerHelmholtz} K. B. Wolf, M. A.  Alonso, G. W. Forbes,   Wigner functions for Helmholtz
wavefields. Journal of the Optical Society of America A \textbf{16} (1999) 2476--2487


\bibitem{SamplingHelmholtz}  P. Gonz\'alez-Casanova, K. B. Wolf,   \textit{Interpolation of solutions to the Helmholtz
 equation}, Numerical Methods of Partial Differential Equations   \textbf{11} (1995) 77--91
 
\bibitem{MeasureingHelmholtz} M.A. Alonso, \textit{Measurement of Helmholtz wave fields}, J. Opt. Soc. Am. A \textbf{17}, 1256-1264 (2000) 


\bibitem{KastrupWigner1} H. A. Kastrup, \textit{Wigner functions for the pair angle and orbital angular momentum}, Phys. Rev. A \textbf{94}, 062113 (2016)

\bibitem{KastrupWigner2} H. A. Kastrup, \textit{Wigner functions for the pair angle and orbital angular momentum: Operators and dynamics}, Phys. Rev. A \textbf{95} 052111 (2017)



\bibitem{MOlmo1} E. Celeghini, M. Gadella, and M. A. del Olmo, \textit{Lie algebra representations and rigged Hilbert spaces: the SO(2) case}, Acta Polytech. \textbf{57}, 379-384 (2017)

\bibitem{MOlmo2} E. Celeghini, M. Gadella, and M. A. del Olmo, \textit{Spherical harmonics and rigged Hilbert spaces}, J. Math.  Physics \textbf{59}, 053502 (2018)



\bibitem{hyperspherical} Zhen-Yi Wen and John Avery, \textit{Some properties of hyperspherical harmonics}, J. Math. Phys. \textbf{26}, 396 (1985)

\bibitem{RieszRepresentationTheorem} W. Rudin, \textit{Real and Complex Analysis}, McGraw-Hill (1966).


 \bibitem{Ali} S. Twareque Ali, R. Roknizadeh  and M. K. Tavassoly, \textit{Representations of coherent states in non-orthogonal bases}, J. Phys. A \textbf{37} (2004) 4407
 

\bibitem{BohmGadella} A. Bohm and M. Gadella, \textit{Dirac Kets, Gamow Vectors and Gelfand Triplets}, Springer Lecture Notes in Physics \textbf{348}, Springer,
Berlin (1989)


 \bibitem{Bochner-Riesz-mean} Shanzhen Lu, Dunyan Yan \textit{Bochner-Riesz Means on Euclidean Spaces}, World Scientific, New Jersey (2013)

\bibitem{PaleyWiener} R. Paley and N. Wiener, \textit{Fourier Transforms in the complex Domain}, Amer. Math. Soc. Colloquium Publ. Ser.  \textbf{19}, Amer. Math. Soc., Providence, Rhode Island (1934)

\bibitem{Hormander} L. H\"ormander, \textit{Linear Partial Differential Operators}, Springer Verlag (1976).
 

%
%
%
%
%
%
%
%
%
%
%
  
%
%
%
%
%
%
%
%
%





\end{thebibliography}
\end{document}